\documentclass{article}%
\usepackage{graphicx}
\usepackage{amsmath}
\usepackage[psamsfonts]{amssymb}
\usepackage[psamsfonts]{amsfonts}
\begin{document}

\title{Elastic Wave Transmission at an Abrupt Junction in a Thin Plate, with
Application to Heat Transport and Vibrations in Mesoscopic Systems}
\author{M.C. Cross and Ron Lifshitz\thanks{Present Address: School of
Physics and Astronomy, Raymond and Beverly Sackler Faculty of
Exact Sciences, Tel Aviv University, Tel Aviv 69978, Israel.}
\\Condensed Matter Physics\\Caltech 114-36, Pasadena CA 91125}
\date{\today              }
\maketitle

\begin{abstract}
The transmission coefficient for vibrational waves crossing an abrupt junction
between two thin elastic plates of different widths is calculated. These
calculations are relevant to ballistic phonon thermal transport at low
temperatures in mesoscopic systems and the $Q$ for vibrations in mesoscopic
oscillators. Complete results are calculated in a simple scalar model of the
elastic waves, and results for long wavelength modes are calculated using the
full elasticity theory calculation. We suggest that thin plate elasticty
theory provide a useful and tractable approximation to the full three
dimensional geometry.

\end{abstract}

%\pacs{62.30.+d, 63.22.+m. 66.70.+f}

\section{Introduction}

The electronic properties of mesoscopic systems have been studied
experimentally and theoretically for many years. More recently the behavior of
other excitations, for example lattice vibrations (phonons) and spin degrees
of freedom, have come under study in these systems. In this paper we present
results relevant to the issues of heat transport by phonons and the
dissipation of vibrational modes in mesoscopic systems.

The interest in heat transport by phonons in mesoscopic systems
arises because for easily fabricated devices the wavelength of a
typical thermal phonon becomes comparable to the dimension of the
thermal pathway at accessible temperatures of order $1K$. Thus
\emph{quantized} thermal transport due to the discrete mode
structure of the thermal pathway should become evident. We
\cite{Angelescu98} and others \cite{Rego98,Blencowe99} showed that
this leads to a natural quantum unit of thermal conductance
$k_{B}^{2}T/h$ similar to the role of $e^{2}/h$ as a quantum of
electrical conductance in one dimensional wires \cite{Landauer57}.
This quantum unit of thermal conductance is predicted
\cite{Rego98} to be clearly observable at low enough temperatures
where only the acoustic $(\omega\rightarrow0$) vibrational modes
are excited in the thermal pathway---the waveguide-like modes with
nonzero frequency cutoffs at long wavelengths are populated with
exponentially small numbers: a universal thermal conductance is
predicted equal to $N_{A}\pi^{2}k_{B}^{2}T/3h$ with $N_{A}$ the
number of acoustic modes ($4$ for a freely suspended beam of
material, corresponding to a longitudinal mode, two bending modes
and a torsional mode). These predictions were recently verified in
beautiful experiments by Schwab et al.\ \cite{Schwab00}.

Vibrational modes in mesoscopic systems are found to have anomalously low $Q$
values, compared to larger systems of the same material
\cite{Mihailovich92,Greywall96,Carr97,Carr99,Harrington00,Mohanty00}. At first
sight, the dissipation might be expected to become smaller as the oscillator
becomes smaller, since defects such as dislocations are eliminated when the
size gets less than a typical defect separation. Thus the observation of lower
values of $Q$ was a surprise. In addition, unexpected dependencies on
temperature \cite{Mohanty00} and magnetic field \cite{Greywall96} have been
observed and remain unexplained.

Both the possibility of observing the universal thermal conductance and
explanations for the $Q$ of small resonators involve the properties of phonon
excitations with a wavelength comparable to the system size. Here we
investigate a particular issue relevant to both these questions, namely the
coupling of vibrational modes across an abrupt junction between two blocks of
the same material but with different dimensions.%

\begin{figure}
[tbh]
\begin{center}
\includegraphics[
width=3.35in
]%
{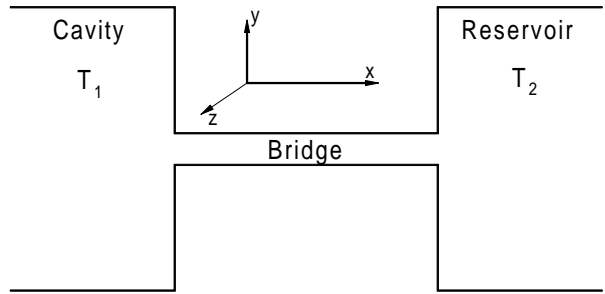}%
\caption{Schematic of possible experimental geometry for the study of thermal
transport and oscillations in mesoscopic systems.}%
\label{Fig_geometry}%
\end{center}
\end{figure}

The geometry typical of a number of experiments is shown in
Fig.~(\ref{Fig_geometry}). The ``bridge'' is made of silicon,
silicon nitride or gallium arsenide, is freely floating, and is of
rectangular cross section. The bridge is connected to two larger
blocks of the same semiconductor. In thermal conduction
experiments the block at one end, called the cavity, is also
freely floating (and is physically supported by four bridges) and
is of the same thickness. The block at the other end provides both
the mechanical support and a thermal reservoir. In recent
experiments \cite{Tighe97} the dimensions were: thickness
$t=200nm$; width $w=300nm$; length $L=5\mu m$. In vibration
experiments the bridge may be supported just at one end (a
cantilever) \cite{Cleland96} or at both ends (a beam)
\cite{Carr99},\cite{Mohanty00}.

An important question in both thermal transport and oscillator damping
experiments is the coupling of the vibrational modes of the bridge to modes in
the supports---how well the energy in a mode in the bridge is transmitted to
the supports, and vice versa. We have previously introduced a simple scalar
model of the elastic waves to study this question in the context of thermal
transport \cite{Angelescu98}. In this paper we give a more realistic
description of the vibrational modes.

First, in section \ref{SecScalar}, we introduce an improved scalar
model, using a better choice for the boundary conditions on the
scalar field that provides a more realistic approximation to the
waves in an elastic medium. The scalar model in its revised form
provides a useful first guide to the expected behavior of the
experimental system, and a simpler environment in which to develop
intuition and methods of theoretical attack. With the scalar model
we perform a complete calculation of the scattering of the waves
at the abrupt junction between the bridge and the supports for all
the modes and at all wave vectors. We use the resulting
transmission coefficients to evaluate the effect of the abrupt
junction on the thermal conductance, particularly the universal
low temperature expression. In addition we introduce a simple
method to calculate the transmission coefficients for the long
wave length acoustic modes, and compare the results with the
general results. In the full elastic calculation we will not be
able to calculate the transmission coefficient for full range of
modes, but will be restricted to this type of long wavelength
calculation.

Secondly we propose that the elasticity theory for a thin plate geometry
provides a useful semiquantitative description of the experimental geometry.
This is described in section \ref{Sec_ThinPlate}. The full elasticity theory
for the modes in the two dimensional, thin plate geometry is sufficiently
tractable that a complete mode spectrum is readily calculated. On the other
hand, a fully three dimensional elasticity theory can only be attacked purely
numerically. The two dimensional theory reproduces many important features of
a fully three dimensional elasticity theory, for example the mixing of bulk
longitudinal and transverse waves by reflection at boundaries, the correct
behavior of the dispersion relation at long wavelength and low frequency,
including the ``bending'' modes with the unusual quadratic dispersion
$\omega\propto k^{2}$ at long wavelengths, and regions of negative dispersion
in the mode spectra. Thus the results should be more informative than the
naive scalar model. The results should be accurate at sufficiently low
temperatures where the modes with structure across the thickness are frozen
out. We use the thin plate model to investigate the mode structure in the
beam, and the coupling of these modes to the supports, also treated as thin
plates of the same thickness.

Finally, in section \ref{Sec_Applications} we apply the results to the issues
of heat transport and oscillations in mesoscopic systems.

%Finally we briefly present for comparison some accessible results for a three
%dimensional geometry of a bridge of rectangular cross section connected to
%supports modeled as infinite three dimensional half-spaces.

\subsection{Heat Transport}%

\begin{figure}
[tbh]
\begin{center}
\includegraphics[
width=2.98in
]%
{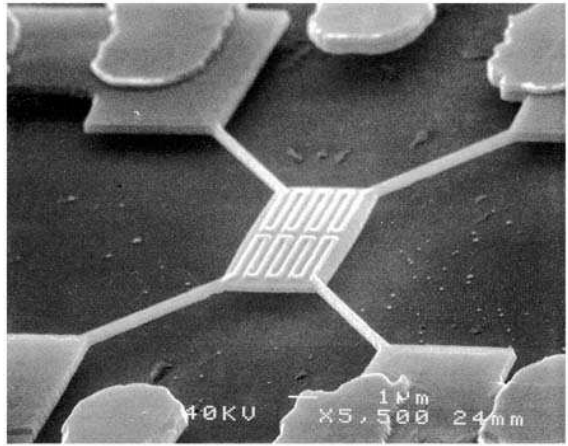}%
\caption{Experimental geometry of Tighe et al. \cite{Tighe97}.}%
\label{Fig_expt}%
\end{center}
\end{figure}
A thermal transport experiment is shown in Fig.~(\ref{Fig_expt}).
Two thermal masses are connected by four thermal pathways of
mesoscopic dimensions in which heat transport by phonons is the
dominant mechanism. One of the thermal masses, which we call the
cavity, is a freely suspended thin block of semiconductor, with
resistive wires on the upper surface to act as heat source and
thermometer. The four bridges act as the thermal pathway to the
outside world, as well as mechanical supports. Conceptually, heat
is added to the cavity by resistive heating, and the resulting
temperature difference from the reservoir is measured, yielding,
for small heating, the thermal conductance of the bridges. In
practice issues such as the thermal contact between the electrons
in the resistive heater and thermometer and the phonons, and other
thermal pathways to the reservoir such as through the electrical
contacts to the resistive heaters, have to be considered. In this
paper we will focus on the ideal situation where the phonon
thermal pathway of the bridge dominates the conductance.

In mesoscopic systems it is easy to cool to temperatures where the
transverse dimensions of the beam are comparable to or smaller
than the typical phonon wavelength $hc/k_{B}T$ where $k_{B}$ is
Boltzmann's constant, $h$ is Planck's constant, $c$ is a typical
speed of sound in the material and $T$ is the temperature. When
this condition is satisfied the discreet mode structure of the
thermal pathway becomes evident---the first level of the
quantization of thermal transport. Angelescu et al.\
\cite{Angelescu98} showed that the thermal conductance takes on a
quasi-universal form, largely independent of the material and mode
structure of the beam, on the assumption that the contact between
the modes in the bridge and the cavity and reservoir can be
considered ideal. An ideal contact implies that the right going
phonon modes in the bridge in Fig.~(\ref{Fig_geometry}) are
populated with a thermal distribution at the temperature of the
cavity, and the left going modes at the temperature of the
reservoir.

In a thermal conductance measurement the cavity and reservoir are maintained
at temperatures $T+\delta T$ and $T$ with temperature difference $\delta T$
small compared to their mean temperature. If we first look at the transport by
the right moving phonons, the energy flux is
\begin{equation}
H^{\left(  +\right)  }=\frac{1}{2\pi}\sum_{m}\int_{0}^{\infty}dk\,v_{gm}%
(k)\,\hslash\omega_{m}(k)\,n(\omega_{m}(k)),
\end{equation}
where $k$ is the wave vector along the bridge, $\omega_{m}(k)$ is the
dispersion relation of the $m$th discrete mode of the bridge, and
$v_{gm}=d\omega_{m}(k)/dk$ is the group velocity. Transforming the integral to
an integral over frequencies yields an expression for the heat transport by
right moving phonons%
\begin{equation}
H^{\left(  +\right)  }=\frac{1}{2\pi}\sum_{m}\int_{\omega_{m}}^{\infty}%
d\omega\,\hslash\omega_{m}(k)\,n(\omega_{m}(k))\,, \label{UnscaledMode}%
\end{equation}
where $\omega_{m}$ is the \emph{cutoff frequency} of the $m$th
mode, i.e. the lowest frequency at which this mode propagates. (We
have assumed the $m$th mode propagates to arbitrarily large
frequencies. If a particular mode only propagates over a finite
band of frequencies, the upper limit of the integral will be
replaced by $\omega_{m}^{\max}$.) \emph{The key simplification in
this result is that the group velocity factor is cancelled by the
density of states in transforming from an integration over wave
numbers to an integration over frequencies.}

For ideal coupling to the reservoirs the distribution function
$n\left( \omega_{m}\left(  k\right)  \right)  $ for the right
moving phonons in Eq.~(\ref{UnscaledMode}) is evaluated as the
Bose distribution at the \emph{cavity} temperature $T+\delta T$.
The thermal conductance is given by subtracting the analogous
expression $H^{\left(  -\right)  }$ for the left moving phonons
given by Eq.~(\ref{UnscaledMode}), but now with the distribution
$n\left(  \omega_{m}\left(  k\right)  \right)  $ given by the
Bose distribution at the \emph{reservoir} temperature $T$%
\begin{equation}
K=\lim_{\delta T\rightarrow0}\frac{H^{(+)}(T+\delta T)-H^{(-)}(T)}{\delta T}.
\end{equation}
Finally, introducing the scaled frequency variable $x=\hbar\omega/k_{B}T$
gives the expression \cite{Angelescu98}%
\begin{equation}
K=\frac{k_{B}^{2}T}{h}\sum_{m}I(\hslash\omega_{m}/k_{B}T)
\label{Eq_Conductance}%
\end{equation}
where $I$ is given by the integral%
\begin{equation}
I(x)=\int_{x}^{\infty}\frac{y^{2}e^{y}}{(e^{y}-1)^{2}}\,dy.
\end{equation}
Equation (\ref{Eq_Conductance}) demonstrates the important result that the
properties of the bridge only enter through the ratio of the mode cutoff
frequencies to the temperature $\hslash\omega_{m}/k_{B}T$. The quantity
$k_{B}^{2}T/h$ plays the role of the quantum unit of thermal conductance,
analogous to the quantum of electrical conductance $e^{2}/h$ for one
dimensional wires. At very low temperatures the contribution to the thermal
conductance by the modes with nonzero cutoff frequency will be exponentially
small leaving a \emph{universal} thermal conductance \cite{Rego98}%
\begin{equation}
K=N_{a}\frac{\pi^{2}k_{B}^{2}T}{3h}%
\end{equation}
where $N_{a}$ is the number of ``acoustic'' modes, i.e. modes with frequency
tending to zero at long wavelengths. Usually this will be \emph{four }for the
beam (two transverse bending modes, one longitudinal compressional mode, and a
torsional mode). Note that there is \emph{no} dependence on the bridge
properties in this expression.

More generally we cannot assume perfect coupling between the modes in the
bridge and the cavity and reservoir. This can be taken into account, following
the Landauer approach to electrical conductance \cite{Landauer57}, through a
transmission coefficient for energy to be transported across the interfaces.
For example for imperfect contact at the cavity bridge interface we would find
a thermal conductance (returning to unscaled quantities in the integral for
clarity)%
\begin{equation}
K=\frac{\hslash^{2}}{k_{B}T^{2}}\sum_{m}\frac{1}{2\pi}\int_{\omega_{m}%
}^{\infty}\mathcal{T}_{m}(\omega)\frac{\omega^{2}e^{\beta\hslash\omega}%
}{(e^{\beta\hslash\omega}-1)^{2}}\,d\omega\label{EqConductivity}%
\end{equation}
where $\mathcal{T}_{m}(\omega)$ is the energy transmission coefficient from
the mode $m$ of the bridge at frequency $\omega$ into the cavity modes.
Imperfect coupling at the bridge-reservoir junction, and elastic scattering
due to imperfections in the bridge can be similarly included through a total
transmission matrix as in the electron case.

A central issue in predicting the thermal conductance is then to
calculate the transmission coefficient $\mathcal{T}_{m}(\omega)$.
This is particularly important in the question of the
observability of the universal conductance at low temperatures,
since the scattering of the long wavelength phonons contributing
to this quantity becomes strong---indeed for the abrupt junction
in Fig.~(\ref{Fig_geometry}),
$\mathcal{T}_{m}(\omega)\rightarrow0$ as $\omega\rightarrow0$ as
we will see below. Although it is feasible in experiment to
``smooth off'' the corners, as indeed was done in the experiment
of Schwab et al.\ \cite{Schwab00}, consideration of the worst case
abrupt junction provides insight into the importance of geometric
scattering.

\subsection{Oscillator Q}

The $Q$ of an oscillator is given by%
\begin{equation}
Q^{-1}=\frac{|\dot{E}|}{\omega E},
\end{equation}
where $\dot{E}$ is the rate of energy loss from the mode at frequency $\omega$
containing energy $E$. If we consider the oscillations of a beam supported by
two supports, or a cantilever with one support, and estimate the energy loss
as the energy transmitted into the supports, we find for the mode $n$%
\begin{equation}
Q_{n}^{-1}\sim\frac{v_{g}}{L\omega}\mathcal{T}_{n}%
\end{equation}
where $v_{g}=d\omega/dk$ is the group velocity of a wave propagating in the
beam, and $L$ is the length of the beam. The exact evaluation of this quantity
depends on the nature of the mode (longitudinal, bending etc.). For the
longitudinal and torsional modes, which have a linear spectrum $\omega=ck$,
the frequency of the fundamental mode in a beam of length $L$ is of order
$c\pi/L$, the group velocity is $v_{g}=c$, and so%
\begin{equation}
Q_{n}^{-1}\sim\frac{\mathcal{T}_{n}}{n\pi}. \label{Q_transmission}%
\end{equation}
For the bending waves with a quadratic spectrum $\omega\propto k^{2}$, the
result is more complicated, but the $Q$ values are similar, and tend to this
form for large $n$, so we will use this expression as a fairly accurate
general estimate.

We see from Eq.~(\ref{Q_transmission}) that good isolation of the mode
$\mathcal{T}_{n}\rightarrow0$ is a criterion for high $Q$. In practice this
expression for the dissipation may be an overestimate, since we are assuming
that all the energy of the mode that enters the support either dissipates
away, or propagates away to large distances so that the energy is not returned
to the oscillations of the beam. If this is not the case, the transmission of
energy into the support is only one part of the problem---we would also have
to consider the behavior of the vibrational energy in the supports as well.

\section{Scalar Model}

\label{SecScalar}

As a simple model of the elasticity problem consider a single scalar field
$\phi$. This might represent, for example, the (scalar) ``displacement'', and
the (vector) ``stress'' $\mathbf{\Sigma}$ would then be proportional to
$\mathbf{\nabla}\phi$. We will suppose a two dimensional domain corresponding
to the thin plate. This leads to a wave equation%
\begin{equation}
\frac{\partial^{2}\phi}{\partial t^{2}}=c^{2}\nabla^{2}\phi
\label{EqScalarWave}%
\end{equation}
with $\nabla^{2}$ the two dimensional Laplacian and $c$ giving the speed of
propagation of the wave. Stress free boundary conditions at the edges are
then
\begin{equation}
\hat{n}\cdot\mathbf{\nabla}\phi=0, \label{EqBC}%
\end{equation}
with $\hat{n}$ the normal to the edge. Note that this Neumann boundary
condition allows the propagation of an acoustic mode ($\omega(k\rightarrow
0)\rightarrow0$) in the bridge as we expect for elastic waves, whereas
Dirichlet boundary conditions do not. An example of an elastic system
described by such a scalar model is a stretched membrane: $\phi$ would then be
the displacement normal to the membrane and $\hat{n}\cdot\mathbf{\Sigma}$ is
the vertical force on a unit line in membrane normal to the direction $\hat
{n}$.

The scalar problem is sufficiently simple that we can calculate the
transmission across an abrupt junction such as the cavity to bridge junction
in full detail. This allows us to gain insight into the more complicated
elastic wave problem, and also allows us to illustrate and test approximation
schemes that will be useful there.

The model Eqs.~(\ref{EqScalarWave}, \ref{EqBC}) was studied by
Angelescu et al.\ \cite{Angelescu98} using the mode matching
method developed by Szafer and Stone \cite{Szafer89} for the
analogous electron wave calculation. However Angelescu et al.\
implicitly used a rather unnatural boundary condition ($\phi=0$)
for the \emph{end} of the cavity at the junction plane (although
Eq.~(\ref{EqBC}) was assumed everywhere else, i.e. on the edges of
the beam and cavity parallel to the propagation direction). We
briefly review this work, and explain how this boundary condition
was introduced, and then treat the more natural case
(Eq.~(\ref{EqBC}) everywhere) using the same methods. This new
treatment actually removes a weak logarithmic divergence found in
the original treatment, and produces results at low frequencies
that are more consistent with the results of the full elasticity
treatment.

\subsection{Model of Angelescu et al.}%

\begin{figure}
[tbh]
\begin{center}
\includegraphics[
width=3.35in
]%
{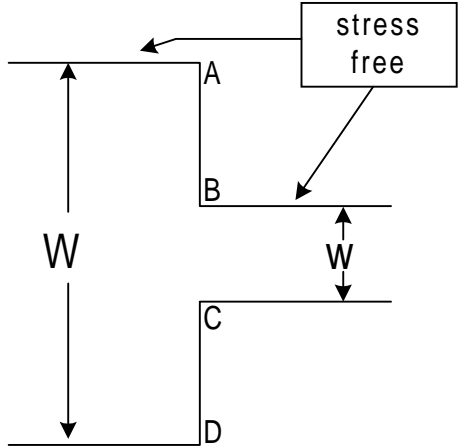}%
\caption{Geometry for the calculation of the transmission coefficient Stress
free boundary conditions are assumed on the edges as shown.
Angelescu et al. used $\phi=0$ boundary conditions on the end $AB$
and $CD$ of the cavity.
A\ better choice is to use stress free conditions here as well.}%
\label{Fig_ScalarSetup}%
\end{center}
\end{figure}
Assume a simple two dimensional geometry consisting of a
rectangular bridge of transverse dimensions $w$ connected to a
rectangular cavity of transverse dimension $W$,
Fig.~(\ref{Fig_ScalarSetup}). In the general three dimensional
case, if the cavity and bridge have the same thickness, there is
no mixing of the $z$ modes, and the problem separates into a set
of two dimensional problems, one for each $z$ mode. Here we will
only consider the lowest mode with no structure across the
$z$-direction, which is the only mode excited at low enough
temperatures. Let $\chi_{\alpha}^{c}(y)$ and $\chi_{m}(y)$ be
orthonormal transverse modes satisfying the stress free boundary
conditions on the edges in the cavity and the bridge,
respectively. (For clarity we will denote cavity mode indices by
Greek letters, and bridge mode indices by Roman letters.)

The solutions to the wave equation take the form%
\begin{equation}
\phi_{m}(x,y,t)=\chi_{m}(y)e^{i(kx-\omega t)}%
\end{equation}
for the bridge, where the frequency of the mode $\omega$ is given by
$\omega^{2}=\omega_{m}^{2}+c^{2}k^{2}$ with $\omega_{m}=m\pi c/w$ the cutoff
frequency of the $m$th bridge mode. The form is similar for the cavity modes
with the cavity width $W$ replacing the bridge width $w$. We will denote the
frequency separation between bridge modes by $\Delta$:
\begin{equation}
\Delta=\omega_{m+1}-\omega_{m}=\pi c/w.
\end{equation}

Consider a phonon incident on the interface from the cavity side ($x<0$), in
the mode $\alpha$ of the cavity, and with longitudinal wave vector $k_{\alpha
}^{\left(  c\right)  }$. The solutions in the cavity and bridge, including the
reflected waves in the cavity and the transmitted waves in the bridge, are:
\begin{equation}%
\begin{array}
[c]{ll}%
\phi^{\left(  c\right)  }=\chi_{\alpha}^{\left(  c\right)  }e^{ik_{\alpha
}^{\left(  c\right)  }x}+\sum_{\beta}r_{\alpha\beta}\chi_{\beta}^{\left(
c\right)  }e^{-ik_{\beta}^{\left(  c\right)  }x} & \text{cavity,}\\
\phi=\sum_{m}t_{\alpha m}\chi_{m}e^{ik_{m}x} & \text{bridge.}%
\end{array}
\label{mode_match}%
\end{equation}
with $r_{\alpha\beta}$ and $t_{\alpha m}$ reflection and transmission
amplitudes to be determined. In the above equations, $k_{m}$ and $k_{\beta
}^{\left(  c\right)  }$ are the wave vectors of the transmitted and reflected
waves respectively, given by the frequency matching condition
\begin{equation}
\omega^{2}=c^{2}k_{\alpha}^{\left(  c\right)  2}+\omega_{\alpha}^{\left(
c\right)  2}=c^{2}k_{m}^{2}+\omega_{m}^{2}=c^{2}k_{\beta}^{\left(  c\right)
2}+\omega_{\beta}^{\left(  c\right)  2}. \label{Energy}%
\end{equation}
Note that the sums over $m$ and $\beta$ in Eq.~(\ref{mode_match}) include
evanescent waves (imaginary $k$ or $k^{\left(  c\right)  }$) although only the
propagating modes will contribute to the energy transport. The field $\phi$
and the longitudinal derivative $\partial\phi/\partial x$ have to be matched
in the medium at $x=0$, which leads to the equations:
\begin{equation}%
\begin{array}
[c]{l}%
\chi_{\alpha}^{\left(  c\right)  }+\sum_{\beta}r_{\alpha\beta}\chi_{\beta
}^{\left(  c\right)  }=\sum_{m}t_{\alpha m}\chi_{m},\\
k_{\alpha}^{\left(  c\right)  }\chi_{\alpha}^{\left(  c\right)  }-\sum_{\beta
}r_{\alpha\beta}k_{\beta}^{\left(  c\right)  }\chi_{\beta}^{\left(  c\right)
}=\sum_{m}t_{\alpha m}k_{m}\chi_{m}%
\end{array}
\label{bound}%
\end{equation}

Equations for the reflection and transmission coefficient are
extracted by integrating Eqs.~(\ref{bound}) multiplied by a
transverse function, $\chi_{m}$ or $\chi_{\alpha}^{\left( c\right)
}$, and using the orthogonality of the functions over the
appropriate domain to extract relationships for the mode
coefficients. We first multiply one of the equations with a
\emph{cavity} mode $\chi_{\beta}^{\left(  c\right)  }$, and
integrate over the cavity width, making use of the orthonormality
relation $\int dy\chi_{\alpha}^{\left( c\right)
}\chi_{\beta}^{\left(  c\right)  }=\delta_{\alpha\beta}$. In this
section we follow Angelescu et al.\ \cite{Angelescu98} and perform
this operation on the \emph{first} equation (i.e. the matching
equation for the field $\phi$). It is at this stage that the
boundary condition on the cavity field at the face $x=0$ for
$|y|>w/2$ is introduced. The replacement in the
integration on the right hand side%
\begin{equation}
\int_{-W/2}^{W/2}dy\chi_{\alpha}^{\left(  c\right)  }\phi(x=0)\Rightarrow
\int_{w/2}^{w/2}dy\chi_{\alpha}^{\left(  c\right)  }\sum_{m}t_{\alpha m}%
\chi_{m}%
\end{equation}
implicitly forces the boundary condition $\phi(x=0)=0$ for $|y|>w/2$.

Multiplying the first equation in (\ref{bound}) by a cavity mode $\chi^{(c)}$,
integrating over the cavity width, and using the orthonormality of the cavity
modes, leads to%
\begin{equation}
r_{\alpha\beta}=-\delta_{\alpha\beta}+\sum_{m}t_{\alpha m}a_{m\beta},
\label{refl}%
\end{equation}
where $a_{m\beta}$ is the overlap of cavity and bridge transverse functions%
\begin{equation}
a_{m\beta}=\int_{-w/2}^{w/2}dy\chi_{\beta}^{\left(  c\right)  }\chi_{m}.
\end{equation}

Equation (\ref{refl}) may now be plugged into the second equation
in Eq.~(\ref{bound}), and the result is:
\begin{equation}
2k_{\alpha}^{\left(  c\right)  }\chi_{\alpha}^{\left(  c\right)  }-\sum
_{m}\sum_{\beta}t_{\alpha m}a_{m\beta}k_{\beta}^{\left(  c\right)  }%
\chi_{\beta}^{\left(  c\right)  }=\sum_{m}t_{\alpha m}k_{m}\chi_{m},
\label{inteq1}%
\end{equation}
which, when integrated with $\chi_{m}$ over the \emph{bridge} width, yields
\begin{equation}
2k_{\alpha}^{\left(  c\right)  }a_{m\alpha}=\sum_{n}A_{nm}t_{\alpha
n}+t_{\alpha m}k_{m}. \label{syst}%
\end{equation}
This is a system of equations that determine $t_{\alpha m}$. In
Eq.~(\ref{syst}) the kernel $A_{mn}$ is given by
\begin{equation}
A_{mn}=\sum_{\beta}a_{m\beta}a_{n\beta}k_{\beta}^{\left(  c\right)  }.
\label{old_a_mn}%
\end{equation}

These equations may be solved for the $t$'s, and then the flux transmission
probability from the wave vector $k_{\alpha}^{\left(  c\right)  }$ state of
cavity mode $\alpha$ to bridge mode $m$ is given by
\begin{equation}
\mathcal{T}_{\alpha m}=|t_{\alpha m}|^{2}\frac{k_{m}}{k_{\alpha}^{\left(
c\right)  }}.
\end{equation}
Now summing over all the cavity modes that are propagating at frequency
$\omega$ leads to the ``transport transmission coefficient'' from the cavity
to the $m$th mode $\mathcal{T}_{m}\left(  \omega\right)  $ (for $\omega
>\omega_{m}$) by%
\begin{equation}
\mathcal{T}_{m}\left(  \omega\right)  =\sum_{\alpha,\omega_{a}^{\left(
c\right)  }<\omega}\mathcal{T}_{\alpha m}=\sum_{\alpha,\omega_{a}^{\left(
c\right)  }<\omega}|t_{\alpha m}|^{2}\frac{k_{m}}{k_{\alpha}^{\left(
c\right)  }}. \label{T_sum}%
\end{equation}
with $k_{m}\left(  \omega\right)  $ and $k_{\alpha}^{\left(  c\right)
}\left(  \omega\right)  $ given by Eq.~(\ref{Energy}). This also gives the
energy transmission coefficient \emph{from} the $m$th bridge mode to the
cavity, by the usual reciprocity arguments.

Equations (\ref{old_a_mn},\ref{T_sum}) involve sums over cavity modes. We may
either evaluate the sums directly for a chosen value of the width ratio $W/w$,
or may take the limit of a large cavity width $W\rightarrow\infty$ when the
sums are replaced by integrals. We calculate the matrix $A_{mn}$ for
$m,n<N_{\max}$ with $N_{\max}$ some upper cutoff for the number of bridge
modes retained and invert the $N_{\max}\times N_{\max}$ matrix system
numerically to find $t_{\alpha m}$ and hence $\mathcal{T}_{m}\left(
\omega\right)  $.

There is a simple approximation \cite{Szafer89} that provides an analytic form
for the solution to Eq.~(\ref{syst}) that is in reasonably good agreement with
the exact solution. The approximation derives from three properties of
$a_{m\alpha}$. Firstly, $a_{m\alpha}=0$ unless $m$ and $\alpha$ have the same
parity. In other words, even modes couple to even modes and odd modes to odd
modes only. Secondly as a function of $\alpha$, $a_{m\alpha}$ is sharply
peaked around $\alpha=mW/w$, the width of the peak being of order $W/w$. And
thirdly, $a_{\alpha m}$ must satisfy the completeness relation
\begin{equation}
\sum_{\alpha}a_{m\alpha}a_{n\alpha}=\delta_{mn}.
\end{equation}
The first two properties permit the key approximation, namely that
$A_{mn}\propto\delta_{mn}$ (since the product of two functions peaked at
different channels $m,n$ is very small and $A_{mn}$ is rigorously zero when
$m$ and $n$ are different parity modes). Then we only need the diagonal part
of $A$%
\begin{equation}
A_{mm}=\sum_{\beta}a_{m\beta}^{2}k_{\beta}^{\left(  c\right)  }, \label{Amn}%
\end{equation}
which is simply a weighted average of the complex wave vector over the narrow
range of reflected cavity modes for which $a_{m\beta}$ is significant. (Note
$\sum_{\beta}a_{m\beta}^{2}=1$ by completeness). In this case Eq.~(\ref{syst})
separates into
\begin{equation}
2k_{\alpha}^{\left(  c\right)  }a_{m\alpha}=A_{mm}t_{\alpha m}+k_{m}t_{\alpha
m}%
\end{equation}
and then
\begin{equation}
t_{\alpha m}=\frac{2k_{\alpha}^{\left(  c\right)  }a_{m\alpha}}{A_{mm}+k_{m}}.
\label{transmission}%
\end{equation}
The flux transmission probability from cavity mode $\alpha$ to bridge mode $m$
is given in this approximation by
\begin{equation}
\mathcal{T}_{\alpha m}=|t_{\alpha m}|^{2}\frac{k_{m}}{k_{\alpha}^{\left(
c\right)  }}\simeq\frac{4k_{\alpha}^{\left(  c\right)  }k_{m}|a_{m\alpha}%
|^{2}}{(k_{m}+K_{m})^{2}+J_{m}^{2}}, \label{tran}%
\end{equation}
where $K_{m}=\operatorname{Re}A_{mm}$ and $J_{m}=\operatorname{Im}A_{mm}$. The
energy transmission coefficient is%
\begin{equation}
\mathcal{T}_{m}\left(  \omega\right)  \simeq\sum_{\alpha,\omega_{a}^{\left(
c\right)  }<\omega}\frac{4k_{\alpha}^{\left(  c\right)  }k_{m}|a_{m\alpha
}|^{2}}{(k_{m}+K_{m})^{2}+J_{m}^{2}}=\frac{4K_{m}k_{m}}{(k_{m}+K_{m}%
)^{2}+J_{m}^{2}}. \label{TmOmega}%
\end{equation}

We now use the explicit form of the transverse modes to evaluate
$\mathcal{T}_{m}\left(  \omega\right)  $. With the boundary condition
$\partial\phi/\partial y=0$ at the $y$ boundaries we have:
\begin{equation}
\chi_{m}(y)=\left\{
\begin{tabular}
[c]{ll}%
$\sqrt{\frac{2}{w}}\cos$($\frac{m\pi y}{w}$) & $m$ even\\
$\sqrt{\frac{2}{w}}\sin(\frac{m\pi y}{w})$ & $m$ odd
\end{tabular}
\ \ \right.
\end{equation}%
\begin{equation}
\chi_{\alpha}^{c}(y)=\left\{
\begin{tabular}
[c]{ll}%
$\sqrt{\frac{2}{W}}\cos$($\frac{\alpha\pi y}{W}$) & $\alpha$ even\\
$\sqrt{\frac{2}{W}}\sin(\frac{\alpha\pi y}{W})$ & $\alpha$ odd
\end{tabular}
\ \ \right.
\end{equation}
with $m$ and $\alpha$ integers, with the special case $\chi_{0}(y)=1/\sqrt{w}$
and $\chi_{0}^{c}(y)=1/\sqrt{W}$. The $a_{m\alpha}$ are easily calculated:%
\begin{align}
a_{m\alpha}  &  =0\quad m,\alpha\text{ not both even or odd }
\label{EqOverlap}\\
a_{m\alpha}  &  =\sqrt{\frac{w}{W}}\left[  \frac{\sin\left(  \frac{\alpha\pi
w}{2W}-\frac{m\pi}{2}\right)  }{\frac{\alpha\pi w}{2W}-\frac{m\pi}{2}%
}+\frac{\sin\left(  \frac{\alpha\pi w}{2W}+\frac{m\pi}{2}\right)
}{\frac{\alpha\pi w}{2W}+\frac{m\pi}{2}}\right]  \text{\quad}m,\alpha\text{
even}\neq0\\
\allowbreak a_{m\alpha}  &  =\sqrt{\frac{w}{W}}\left[  \frac{\sin\left(
\frac{\alpha\pi w}{2W}-\frac{m\pi}{2}\right)  }{\frac{\alpha\pi w}%
{2W}-\frac{m\pi}{2}}-\frac{\sin\left(  \frac{\alpha\pi w}{2W}+\frac{m\pi}%
{2}\right)  }{\frac{\alpha\pi w}{2W}+\frac{m\pi}{2}}\right]  \text{\quad
}m,\alpha\text{ odd}\neq0\\
a_{0\alpha}  &  =\sqrt{\frac{2w}{W}}\left[  \frac{\sin\left(  \frac{\alpha\pi
w}{2W}\right)  }{\frac{\alpha\pi w}{2W}}\right]  \quad\alpha\text{ even}%
\neq0\\
a_{m0}  &  =0\quad m\neq0\\
a_{00}  &  =1
\end{align}
For large $m,\alpha$ (both even or both odd) we can approximate%
\begin{equation}
a_{m\alpha}\simeq\sqrt{\frac{w}{W}}\frac{\sin\left(  \frac{\alpha\pi w}%
{2W}-\frac{m\pi}{2}\right)  }{\frac{\alpha\pi w}{2W}-\frac{m\pi}{2}}.
\label{aApprox}%
\end{equation}
The $a_{m\alpha}$ are indeed sharply peaked as a function of cavity mode
number $\alpha$. The large $m$ approximation is essentially identical to the
result in the electronic case \cite{Szafer89}. However the small, second term
in the braces in Eq.~(\ref{EqOverlap}) ignored in this approximation appears
with the opposite sign in our application. This turns out to render the sum
over the cavity modes $\beta$ appearing in Eq.~(\ref{old_a_mn}) weakly
(logarithmically) divergent for large $\beta$. (Note that $k_{\beta}^{(c)}$ is
imaginary here, so this divergent contribution is to $\operatorname{Im}A_{mn}$
and to the component $J_{m}$ of the diagonal terms.) We must impose some upper
cutoff to the sum to achieve finite results. Physically we might suppose such
a cutoff may come from the breakdown of the sharp corner approximation at
short enough scales, or ultimately, in a perfectly fabricated mesoscopic
system, from the atomic nature of the material leading to a finite number of
modes.%

\begin{figure}
[tbh]
\begin{center}
\includegraphics[
height=4.0075in,
width=4.9882in
]%
{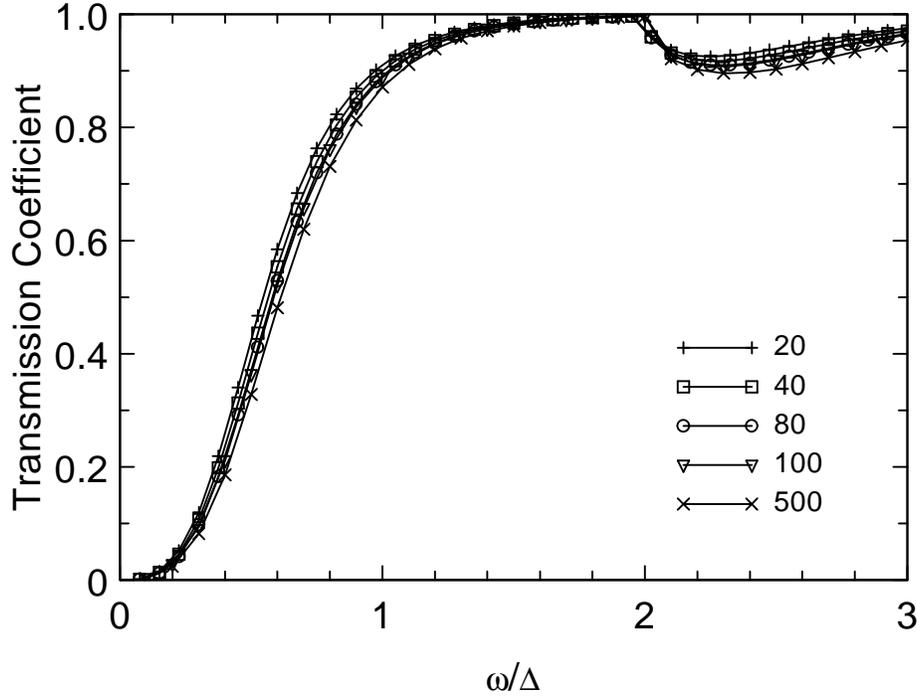}%
\caption{Transmission coefficient coupling the lowest bridge mode to the
cavity modes as a function of the reduced frequency of the mode $\omega
/\Delta$ with $\Delta$ the splitting between bridge modes at zero wave vector.
Curves for cavity wave number cutoff equal to $20,$ $40,$ $80,$ $100,$ $500$
times $\pi w^{-1}$ show the weak dependence on this cutoff.}%
\label{Fig_old_cutoff}%
\end{center}
\end{figure}
Results for the transmission coefficient of the lowest bridge mode
$\mathcal{T}_{0}(\omega)$ are shown in
Fig.~(\ref{Fig_old_cutoff}). There is only a weak dependence on
the cavity sum cutoff. The dependence at small frequency fits well
the expected \cite{Landau86} cubic dependence
$\mathcal{T}_{0}(\omega)\propto\omega^{3}$. The results shown were
calculated for the case of an infinitely wide cavity (sums over
cavity modes replaced by integrals). Results for finite widths
(e.g. $W/w=20$) are very similar. The first 11 bridge modes (6
even modes) were retained in the matrix inversion for the results
shown: increasing this number did not change the results
significantly showing that $A_{mn}$ indeed decreases rapidly for
increasing $|m-n|$. Note that $\mathcal{T}_{0}(\omega)$ rapidly
approaches unity as the frequency grows. There is a small
($\lesssim10\%$) decrease at the frequency $\omega=2\Delta$ where
the \emph{second }even bridge mode becomes propagating, and
similar features of reducing size occur at subsequent integral
multiples of $2\Delta$. There is no coupling between even and odd
modes for the
symmetric geometry used.%

\begin{figure}
[tbh]
\begin{center}
\includegraphics[
height=4.0075in,
width=4.9882in
]%
{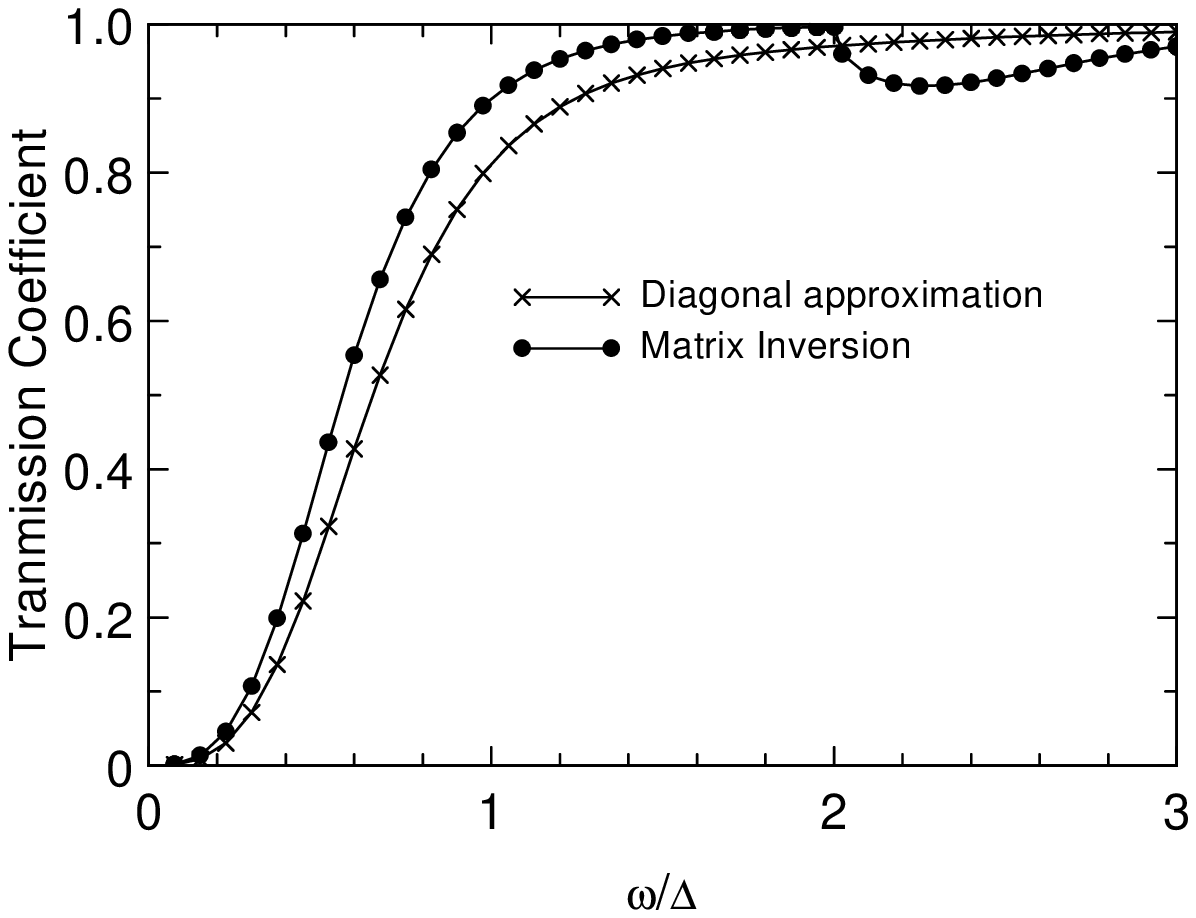}%
\caption{Tranmssion coefficient for the lowest bridge mode in the Angelescu et
al.\ scalar model calculated using the (exact) matrix
diagonalization (points) and the diagonal approximation (crosses)
for a cavity mode wave vector cutoff $40\pi w^{-1}$. The diagonal
approximation shows a somewhat stronger dependence on this cutoff
than shown by the exact results, so that this
comparison will vary as we change the cutoff assumption.}%
\label{Fig_old_e-d}%
\end{center}
\end{figure}
The comparison between the results from the full matrix inversion
and the diagonal approximation is shown in
Fig.~(\ref{Fig_old_e-d}). The diagonal approximation gives results
good to about $10\%$. However this comparison depends on the
cavity wave number cutoff, since the diagonal approximation
depends rather more strongly on this parameter than for the full
results shown in Fig.~(\ref{Fig_old_cutoff}). Note that the
feature at $\omega=2\Delta$ due to the interaction between
different bridge modes is absent in the diagonal approximation.

\subsection{Stress Free Ends}

\subsubsection{Mode matching calculation\label{Subsec_ModeMatching}}

We now redo the scalar analysis, enforcing the boundary condition $\hat
{n}\cdot\mathbf{\nabla}\phi=0$ on the end of the cavity $x=0$, $|y|>w/2$. This
corresponds to a stress free boundary everywhere.

The analysis proceeds as before up to Eq.~(\ref{mode_match}). But now we first
multiply the \emph{second} equation (for the continuity of $\partial
\phi/\partial x$) by $\chi_{\beta}^{\left(  c\right)  }$, and integrate over
the cavity width. This enforces the boundary condition on the cavity face%
\begin{equation}
\hat{n}.\mathbf{\nabla}\phi=0\text{ for }x=0\text{ and }|y|>w/2.
\end{equation}
The orthogonality of the $\chi_{\beta}^{\left(  c\right)  }$gives%
\begin{equation}
r_{\alpha\beta}k_{\beta}^{(c)}=k_{\alpha}^{(c)}\delta_{\alpha\beta}-\sum
_{m}t_{\alpha m}k_{m}a_{m\beta}.
\end{equation}
Use this equation to eliminate $r_{\alpha\beta}$ from the first of
Eq.~(\ref{mode_match})%
\begin{equation}
2\chi_{\alpha}^{\left(  c\right)  }-\sum_{n}\sum_{\beta}t_{\alpha n}%
(k_{n}/k_{\beta}^{\left(  c\right)  })a_{n\beta}\chi_{\beta}^{\left(
c\right)  }=\sum_{m}t_{\alpha n}\chi_{n}%
\end{equation}
and integrate with $\chi_{m}$ over the bridge width to yield%
\begin{equation}
2a_{m\alpha}=\sum_{n}\bar{A}_{mn}k_{n}t_{\alpha n}+t_{\alpha m} \label{t_bar}%
\end{equation}
where%
\begin{equation}
\bar{A}_{mn}=\sum_{\alpha}a_{m\alpha}a_{n\alpha}/k_{\alpha}^{(c)}.
\label{Abar}%
\end{equation}
Again we can solve the equation for $t_{\alpha m}$ Eq.~(\ref{t_bar})
numerically or by using the diagonal approximation for $\bar{A}_{mn}$. It is
easily seen that the extra inverse powers of $k_{\alpha}^{(c)}$ in
$\bar{A}_{mn}$ render the sum over $\alpha$ convergent, unlike the case for
$A_{mn\text{.}}$The energy transmission coefficient from the $m$th bridge mode
remains given by Eq.~(\ref{T_sum}). The diagonal approximation $\bar{A}_{mn}%
\simeq\bar{A}_{mm}\delta_{mn}$ now leads to%
\begin{equation}
\mathcal{T}_{m}\left(  \omega\right)  \simeq\frac{4\bar{K}_{m}k_{m}}%
{(k_{m}\bar{K}_{m}+1)^{2}+\bar{J}_{m}^{2}k_{m}^{2}}%
\end{equation}
with $\bar{J}_{m}=\operatorname{Re}\bar{A}_{mm}$ and $\bar{K}_{m}%
=\operatorname{Im}\bar{A}_{mm}$.%

\begin{figure}
[tbh]
\begin{center}
\includegraphics[
height=4.0075in,
width=4.9882in
]%
{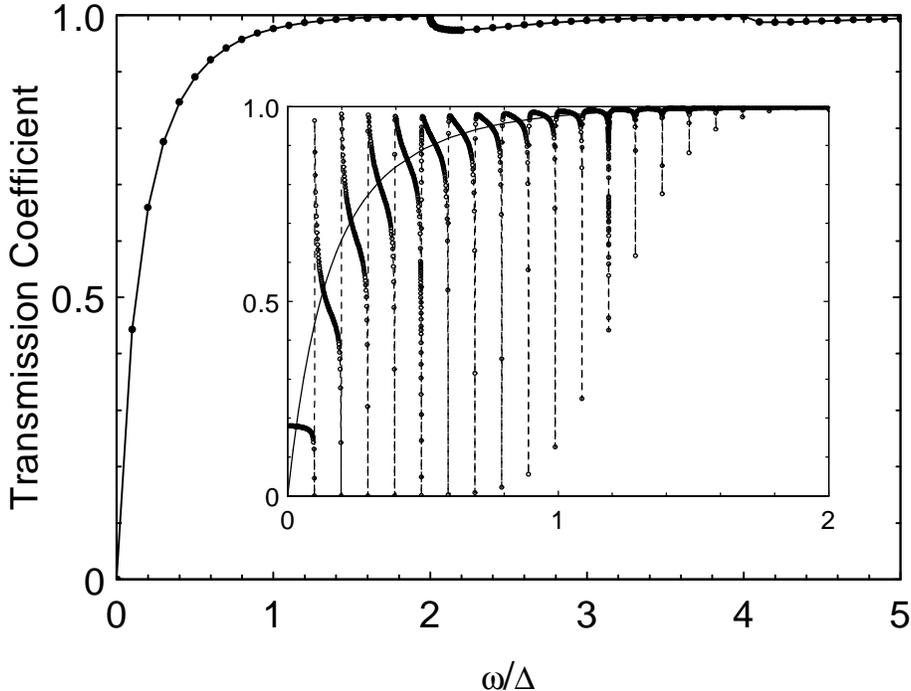}%
\caption{Transmission coefficient coupling the lowest bridge mode to the
cavity modes as a function of the reduced frequency $\omega/\Delta$ in the
scalar model with stress free boundaries. The main graph is for an infinite
cavity width. The inset shows the comparison with results for a finite cavity
width ($W=$ $20.217w$).}%
\label{Fig_new}%
\end{center}
\end{figure}
The result for $\mathcal{T}_{0}\left(  \omega\right)  $ for the
lowest bridge mode is shown in Fig.~(\ref{Fig_new}) taking the
limit of an infinite cavity width (the sums in Eq.~(\ref{Abar})
etc. evaluated as integrals). Again the transmission coefficient
grows rapidly, approaching close to unity as the frequency grows,
for example reaching $0.9$ by about $\omega\sim0.5\Delta$. Note
the important difference from the previous scalar model that the
low frequency asymptotic behavior is \emph{linear},
$\mathcal{T}_{0}\left( \omega\right)  \propto\omega$, rather than
cubic as obtained there, and so $\mathcal{T}_{0}\left(
\omega\right)  $ approaches unity more rapidly than anticipated in
that work. A small reduction in $\mathcal{T}_{0}\left(
\omega\right)  $ (by about $3\%$) near $\omega=2\Delta$ and by
smaller amounts at higher integral multiples of this frequency are
apparent. An analysis of the curve in this region shows a square
root dependence on $\omega-2\Delta$, corresponding to a coupling
in the full matrix calculation to the second bridge mode, and to
the square root growth in $\mathcal{T}_{2}\left( \omega\right)  $
that occurs here. The diagonal approximation (not shown) gives
results that are indistinguishable on the figure for $0\leq\omega
\leq2\Delta$, but the small decrease above $\omega=2\Delta$ does
not appear in this approximation.

The inset in Fig.~(\ref{Fig_new}) shows the comparison with results for a
\emph{finite} cavity width. The results are quite surprising, with
resonance-like features occurring whenever the frequency passes through the
cutoff frequency of a \emph{cavity} mode. This can be traced to the inverse
power of $k^{(c)}$ occurring in Eq.~(\ref{T_sum}) and Eq.~(\ref{Abar}). Since
$k^{(c)}$ goes to zero as $\sqrt{\omega-\omega_{n}^{(c)}}$ these singularities
are integrable, and the features disappear in the limit of infinite cavity
width. Smoothing over the features (e.g. taking the average over bins between
successive $\omega_{n}^{(c)}$) gives points that follow the smooth curve for
the infinite cavity width closely. Integrating the effect of the transmission
coefficient in the thermal conductance over the thermal factors will
effectively perform this smoothing, so that the features will not be apparent
in the thermal measurements. The sharp features are presumably also smoothed
out if the junction is not perfectly abrupt.%

\begin{figure}
[tbh]
\begin{center}
\includegraphics[
height=4.0075in,
width=4.9882in
]%
{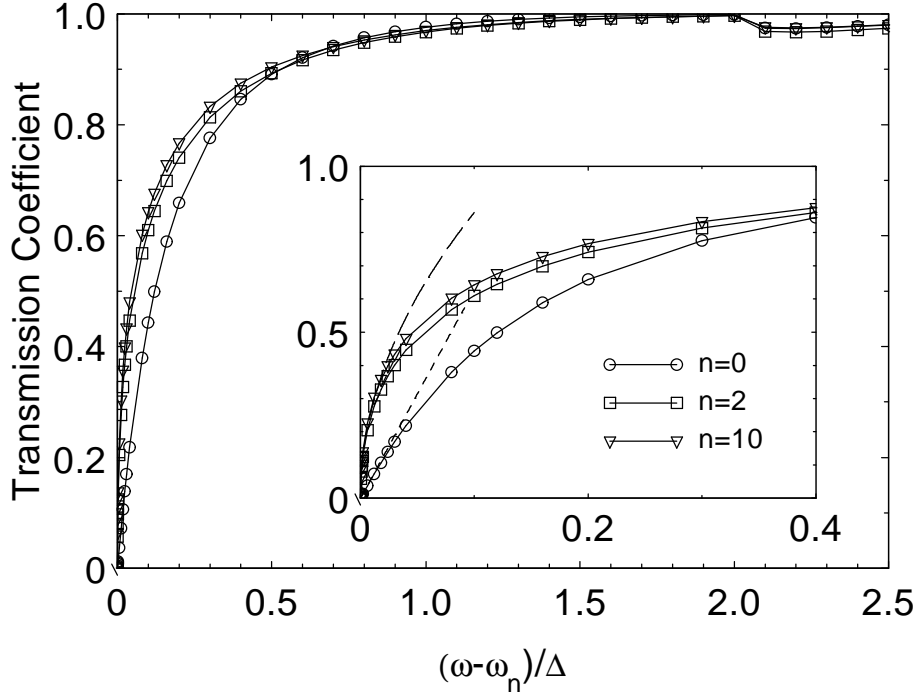}%
\caption{Values of the transmission coefficients from the $n$th bridge mode to
the cavity as a function of $(\omega-\omega_{n})/\Delta$ with $\omega_{n}$ the
cutoff frequency for the $n$th mode and $\Delta$ the spacing between mode
cutoff frequencies $\Delta=\omega_{n+1}-\omega_{n}$. The long-dashed line
shows the square root dependence near cutoff for the $n\neq0$ modes. The
short-dashed line is the linear dependence $T_{0}(\omega)\simeq2\pi
\omega/\Delta$ expected for the lowest mode.}%
\label{new_many}%
\end{center}
\end{figure}
Results for $\mathcal{T}_{n}\left(  \omega\right)  $ for other
values of $n$
are shown in Fig.~(\ref{new_many}).%

\begin{figure}
[tbh]
\begin{center}
\includegraphics[
height=4.0577in,
width=5.0427in
]%
{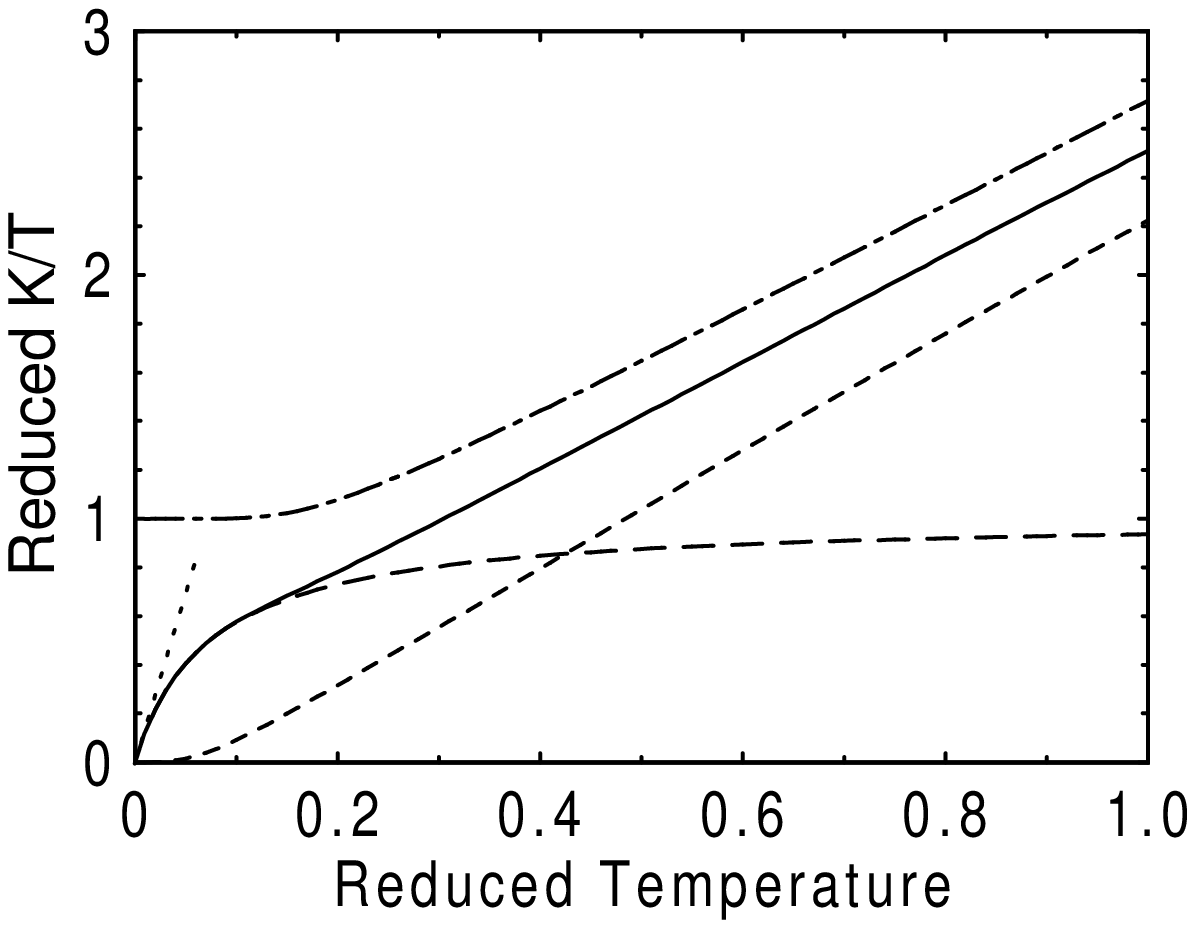}%
\caption{Thermal conductivity divided by temperature reduced by the zero
temperature universal value $\pi^{2}k_{B}^{2}/3h$ as a function of
the reduced temperature $k_{B}T/\hbar\Delta$: solid curve - full
calculation including the transmission losses due to the abrupt
junction for stress free face; long-dashed curve - contribution
from the lowest (acoustic) mode, showing the reduction at low
temperatures due to the scattering at the abrup junction;
dash-dotted curve - ideal result from all modes with full
transmission; dotted line - low temperature asymptotic slope
predicted from the low frequency behavior of the transmission
coefficient. The short-dashed curves shows the result including
scattering of the calculation of Angelescu et al.\
\cite{Angelescu98}.}%
\label{Fig_K_Tnew}%
\end{center}
\end{figure}
It is now straightforward to calculate the thermal conductivity
using Eq.~(\ref{EqConductivity}). We focus on the low temperature
limit where the universal result for $K/T$ applies in the ideal
limit. With no reduction in the thermal transport due to
scattering at the junction the plot of $K/T$ as a function of
temperature develops a plateau at low temperatures at the
universal value $\pi^{2}k_{B}^{2}/3h$ (see
Fig.~(\ref{Fig_K_Tnew}). In this regime the conductivity is
dominated by the acoustic mode with $\omega \rightarrow0$ as
$k\rightarrow0$. Scattering at the junction reduces the
transmission of the small $\omega,k$ waves, so that the value of
$K/T$ is reduced from the no-scattering value. As can be seen from
Fig.~(\ref{Fig_inplane_2}) this reduction begins to occur as the
temperature is lowered at about the same temperature at which the
plateau in the ideal case begins to develop. This suggests that
the plateau at the universal value of $K/T$ will not be well
developed for the abrupt junction. The full calculation using
Eq.~(\ref{EqConductivity})\ and the reduced transmission
coefficients of all the modes (solid curve in
Fig.~(\ref{Fig_inplane_2})) shows that this is the
case---including the effects of scattering at the abrupt junction
yields a $K/T$ curve that tends smoothly to zero as
$T\rightarrow0$. This result clearly demonstrates the importance
of using smooth junctions between the bridge and the reservoirs,
such as was done in the experiments of Schwab et al.\
\cite{Schwab00}, if the universal value of $K/T$ is to be
apparent.

\subsubsection{Long-Wavelength Calculation\label{secMPscalar}}

Although we have performed the full calculation of the transmission
coefficient for the scalar case, this will not be possible for the elasticity
theory calculation. We therefore introduce a simple analytic technique for
establishing the low frequency limit of $\mathcal{T}_{0}\left(  \omega\right)
$ that can be extended to the full elasticity description.

The approach relies on the poor transmission $\mathcal{T}_{0}\left(
\omega\right)  \rightarrow0$ for $\omega\rightarrow0$ (long wavelength waves
are strongly affected by the abrupt junction) to treat the transmission
perturbatively. First the bridge mode is calculated assuming perfect
reflection i.e. isolated from the supports. A simple analysis shows that the
appropriate boundary condition on the end of the bridge is zero
\emph{displacement} (rather than zero stress). If we now ``reconnect'' the
bridge to the supports, the stress fields at the end face of the bridge act as
radiation sources of waves into the cavity. The total power in these waves for
unit incident amplitude in the bridge gives us the transmission coefficient.
The total power radiated may be readily calculated by integrating the product
of the stress source and resulting velocity across the end face of the bridge.

To establish the appropriate boundary conditions in the $\mathcal{T}%
_{0}\left(  \omega\right)  \rightarrow0$ limit we first consider a
\emph{finite} cavity width i.e. an abrupt junction between a bridge of width
$w$ for $x<0$ and a cavity of finite width $W$ $\gg w$ for $x>0$ in the limit
$\omega,k\rightarrow0$. (It is simplest here to consider the transmission from
bridge to cavity, and we have reversed the sense of the $x$ coordinate
compared to Sec. \ref{Subsec_ModeMatching}.) For $x<0$ we consider an incident
wave of unit amplitude and a reflected wave of amplitude $r$%
\begin{equation}
\phi=e^{ikx-i\omega t}+re^{-ikx-i\omega t},\quad x<0. \label{scalar_left}%
\end{equation}
For $x>0$ there is only the transmitted wave%
\begin{equation}
\phi=te^{ikx-i\omega t},\quad x>0. \label{scalar_right}%
\end{equation}
Here the wave numbers $k$ are fixed by the dispersion relation $\omega=ck$,
which is the same on both sides of the junction for the acoustic mode.

The reflection and transmission amplitudes can be calculated by a simple
matching at $x=0.$ (This is equivalent to a wave impedance calculation).
Matching the displacement field $\phi$ gives%
\begin{equation}
1+r=t
\end{equation}
and matching the total force gives%
\begin{equation}
w(1-r)=Wt\text{.}%
\end{equation}
Note that the force (the integrated stress) is conserved because
there are no additional stresses on the cavity face for $|y|>w/2$:
this matching would not be appropriate for the boundary condition
used by Angelescu et al.\ \cite{Angelescu98}.

Thus we find%
\begin{equation}
r=-\frac{W-w}{W+w},\quad t=\frac{2w}{w+W}.
\end{equation}
These expressions are good for $kW,kw\ll1$. In this limit the matching
conditions can be applied outside of the region close to the junction where
the fields are perturbed from their asymptotic forms Eqs.~(\ref{scalar_left}%
,\ref{scalar_right}) but before the exponential phase factors of the wave
propagation have significantly changed. For $W\gg w$ these expressions reduce
to $r\simeq-1$, $t\simeq0$, i.e. strong mismatch and almost perfect reflection
with a sign change. Note that at $x=0$ this implies $\phi\simeq0$, i.e. zero
displacement boundary conditions, together with $\partial\phi/\partial
x\simeq2ik$ for unit incident amplitude. We now use this result as the basis
of the calculation of the transmission for $k$ small but nonzero and the limit
$W\rightarrow\infty$.

The transmission at $k\neq0$ is calculated as a radiation problem, namely via
the power radiated by the end of the bridge into the cavity. The zeroth order
approximation for the solution in the bridge is the perfect reflection result
calculated above, giving the stress radiation source on the wall of the
cavity
\begin{equation}
s(y)=\left\{
\begin{tabular}
[c]{ll}%
$\partial\phi/\partial x\simeq2ike^{-i\omega t}$ & for $|y|<w/2$\\
$0$ & for $|y|>w/2$%
\end{tabular}
\ \right.
\end{equation}
where the second line is just the stress free boundary condition%
\begin{equation}
\partial\phi/\partial x=0\quad\text{for }x=0,|y|>w/2.
\end{equation}

The problem of the radiation due to a stress source on the boundary of an
elastic half space is known as the Lamb problem in elasticity theory, and has
been considered by many authors, for example reference \cite{Miller54}. The
radiation field can be calculated by standard Fourier transform techniques.
The solution in the cavity for $x>0$ can be written%
\begin{equation}
\phi=e^{-i\omega t}\frac{1}{2\pi}\int_{-\infty}^{\infty}\tilde{\phi}%
(\zeta)e^{iqx}e^{i\zeta y}d\zeta
\end{equation}
with%
\begin{equation}
q=\left\{
\begin{tabular}
[c]{lll}%
$\sqrt{k^{2}-\zeta^{2}}$ & for & $|\zeta|\leq k$\\
$i\sqrt{k^{2}-\zeta^{2}}$ & for & $|\zeta|>k$%
\end{tabular}
\ \ \ \right.
\end{equation}
corresponding to propagation or exponential decay away from the interface. The
transform $\tilde{\phi}(\zeta)$ is given by matching to the known
$\partial\phi/\partial x$ at $x=0$ yielding%
\begin{equation}
iq\tilde{\phi}(\zeta)=\int_{-\infty}^{\infty}s(y)e^{-i\zeta y}dy.
\end{equation}
The power radiated is the product of the stress $\partial\phi/\partial x$ and
the velocity $\partial\phi/\partial t$ across the radiation source%
\begin{equation}
P=\left\langle \int_{-w/2}^{w/2}\frac{\partial\phi}{\partial t}\frac{\partial
\phi}{\partial x}dy\right\rangle _{x=0}%
\end{equation}
where the $<>$ denotes the time average. For the fields varying as
$e^{-i\omega t}$ this gives%
\begin{equation}
P=\frac{1}{2}\left.  \operatorname{Re}\left[  -i\omega\int_{-w/2}^{w/2}%
\phi(y)s^{\ast}(y)\,dy\right]  \right|  _{x=0}.
\end{equation}
Inserting the Fourier expression for $\phi(x=0)$%
\begin{align}
P  &  =\frac{\omega}{4\pi}\operatorname{Im}\int_{-\infty}^{\infty}d\zeta
\tilde{\phi}(\zeta)\int_{-\infty}^{\infty}dy\,s^{\ast}(y)e^{i\zeta y}\\
&  =\frac{\omega}{4\pi}\operatorname{Im}\int_{-\infty}^{\infty}d\zeta
\frac{1}{iq}\left|  \int_{-\infty}^{\infty}s(y)e^{-i\zeta y}dy\right|  ^{2}%
\end{align}
Since the transverse wave vector $\zeta$ is limited by $\zeta<k$ for the
integrand to be imaginary (corresponding to $q$ real, i.e. propagating waves
in the cavity), and $s(y)$ is nonzero only for $|y|<w/2$ an expansion in
$kw\ll1$ is given by the expansion%
\begin{equation}
\int_{-\infty}^{\infty}s(y)e^{-i\zeta y}dy=s_{0}-i\zeta s_{1}+\cdots
\end{equation}
where $s_{i}$ are successive moments of the source%
\begin{align}
s_{0}  &  =\int s(y)dy,\\
s_{1}  &  =\int s(y)ydy.
\end{align}
We will need only the leading order term, i.e. $s_{0}$ if the source is parity
symmetric in $y$, $s_{1}$ if the source is antisymmetric. In particular for
the present case $s_{0}=2ikw$ so that%
\begin{equation}
P=\frac{\omega}{2\pi}\int_{0}^{k}d\zeta\frac{4k^{2}}{\sqrt{k^{2}-\zeta^{2}}}%
\end{equation}
Normalizing by the power in the incident wave $\frac{1}{2}w\omega k$ gives the
transmission coefficient%
\begin{equation}
\mathcal{T}_{0}(\omega)=\frac{4kw}{\pi}\int_{0}^{k}\frac{1}{\sqrt{k^{2}%
-\zeta^{2}}}d\zeta
\end{equation}

The integral expression for $\mathcal{T}_{0}$ can be easily
understood physically as integrating over the power radiated into
the continuum of waves, traveling at the wave propagation speed
$c$, and propagating at all angles into the half space. For the
scalar model there is a single type of propagating wave. With the
full elasticity theory we will see a similar result, but with a
number of propagating waves contributing to the power
radiated.

Performing the integral gives%
\begin{equation}
\mathcal{T}_{0}(\omega)=2kw=2\omega w/c\quad\text{for\quad}kw\ll1.
\end{equation}
This result confirms the \emph{linear} dependence on $\omega$ for small
$\omega$, as shown in Fig.~(\ref{new_many}).

\section{Thin Plate Theory}

\label{Sec_ThinPlate}

A useful model of the mesoscopic geometry, that is more tractable than a fully
three dimensional elasticity calculation, is to assume a thin plate geometry.
Thus we take the elastic structure to be carved from a thin plate of uniform
thickness $d$, which is taken to be small with respect to the other dimensions
and also with respect to the wavelength of the elastic waves. In this model
the mode frequency cutoffs at $k=0$, important in the thermal conductance, can
be readily calculated (for most of the modes they are given by simple analytic
expressions). In addition, although the mode structure is quite complicated,
involving the mixing of longitudinal and transverse components by reflection
at the edges, explicit expressions for the transverse structure of the modes
in terms of a finite sum of sinusoidal functions can be written down. Compare
this with a full three dimensional analysis, where a finite dimensional
representation of the modes is not possible. In addition to the scattering at
an abrupt junction pursued here, the thin plate limit will permit an analysis
of phenomena such as the scattering of the waves of surface and bulk
imperfections, an issue we hope to pursue in future work.

We first review the general elastic theory for waves in a thin plate, and
confirm the expected dispersion relations for the acoustic modes of a
rectangular beam (i.e. the ``bridge'') in the long wavelength limit, and then
use the equations to study the coupling of long wavelength modes across the
abrupt bridge-support junctions.

\subsection{Review of Elastic Theory and Modes}

The elasticity of an isotropic solid is summarized by the relationship between
the stress tensor $\mathbf{T}$ and the strain tensor%
\begin{equation}
T_{ij}=-K\Theta\delta_{ij}-2\mu\Sigma_{ij}.
\end{equation}
Here $\Theta$ is the dilation and $\mathbf{\Sigma}$ is the shear strain%
\begin{subequations}
\begin{align}
\Theta &  =\frac{\partial u_{x}}{\partial x}+\frac{\partial u_{y}}{\partial
y}+\frac{\partial u_{z}}{\partial z}\\
\Sigma_{ij}  &  =\frac{1}{2}\left(  \frac{\partial u_{i}}{\partial x_{j}%
}+\frac{\partial u_{j}}{\partial x_{ij}}\right)  -\frac{1}{3}\Theta\delta_{ij}%
\end{align}
with $\mathbf{u}(\mathbf{x})$ the displacement and $K$ and $\mu$
elastic constants.

For a thin plate of thickness $d$ in the $xy$ plane, linear elasticity theory
can be separated into equations for the normal displacement $u_{z}=w(x,y)$ of
the plate and for the in-plane displacements averaged over the depth
$\mathbf{u}(x,y)=(u,v)$ with $u=\left\langle u_{x}(x,y,z)\right\rangle _{z}$
and $v=\left\langle u_{y}(x,y,z)\right\rangle _{z}$, all functions of just two
spatial variables. This is done by assuming, for in plane wave vectors
$\mathbf{k}$ such that $kd\ll1$, that the stresses in the vertical direction
$\Sigma_{zj}$, which must be zero at the nearby stress free top and bottom
surfaces, may be put to zero everywhere. This allows the variation of the
strains across the thickness of the plates to be eliminated in terms of the
variables $u,v,w$. For example setting $T_{zz}$ to zero gives%
\end{subequations}
\begin{equation}
\frac{\partial u_{z}}{\partial z}=-\frac{K-\frac{2}{3}\mu}{K+\frac{2}{3}\mu
}\left(  \frac{\partial u}{\partial x}+\frac{\partial v}{\partial y}\right)  .
\label{eliminate_uz}%
\end{equation}
Thus the modes separate into modes with in-plane polarizations, and modes with
polarizations normal to the plane (flexural modes). The full development can
be found in any standard text on elasticity, for example Landau and Lifshitz
\cite{Landau86} or Graff \cite{Graff91}. Here we collect the main results.

\subsubsection{In-plane polarization}

Using relationships such as Eq.~(\ref{eliminate_uz}) leads to an effective
\emph{two} dimensional elasticity theory for displacements in the plane,
summarized by the stress-strain relationship%
\begin{equation}
T_{ij}^{(2)}=-\bar{K}\Theta^{(2)}\delta_{ij}-2\bar{\mu}\Sigma_{ij}^{(2)}.
\label{stress-strain2}%
\end{equation}
where the indices $i,j$ now run only over $x$ and $y$, $\Theta^{(2)}$ and
$\mathbf{\Sigma}^{(2)}$ are the two dimensional dilation and shear strain
tensor%
\begin{subequations}
\begin{align}
\Theta^{(2)}  &  =\frac{\partial u_{x}}{\partial x}+\frac{\partial u_{y}%
}{\partial y}\\
\Sigma_{ij}^{(2)}  &  =\frac{1}{2}\left(  \frac{\partial u_{i}}{\partial
x_{j}}+\frac{\partial u_{j}}{\partial x_{ij}}\right)  -\frac{1}{2}\Theta
^{(2)}\delta_{ij}%
\end{align}
and the effective two dimensional elastic constants are%
\end{subequations}
\begin{subequations}
\begin{align}
\bar{\mu}  &  =\mu\\
\bar{K}  &  =\frac{3K\mu}{K+\frac{4}{3}\mu}.
\end{align}
The elastic waves with in-plane polarization are then given by the equation of
motion%
\end{subequations}
\begin{align}
\rho\frac{\partial^{2}\mathbf{u}}{\partial t^{2}}  &  =-\nabla_{\perp}%
\cdot\mathbf{T}^{(2)}\\
&  =\bar{K}\nabla_{\perp}(\nabla_{\perp}\cdot\mathbf{u})+\bar{\mu}%
\nabla_{\perp}^{2}\mathbf{u}.
\end{align}
with $\mathbf{\nabla}_{\perp}$ the horizontal gradient operator. In a
horizontally infinite sheet there are longitudinal and transverse waves with
speeds%
\begin{subequations}
\begin{align}
c_{L}^{2}  &  =\frac{\bar{K}+\bar{\mu}}{\rho}=\frac{4\mu}{\rho}%
\frac{(K+\frac{1}{3}\mu)}{(K+\frac{4}{3}\mu)},\\
c_{T}^{2}  &  =\frac{\bar{\mu}}{\rho}=\frac{\mu}{\rho}.
\end{align}
Alternatively, introducing the Young's modulus $E$ and Poisson ration $\sigma$
(with $-1\leq\sigma\leq1/2$) so that%
\end{subequations}
\begin{equation}
K=\frac{E}{3(1-2\sigma)},\quad\mu=\frac{E}{2(1+\sigma)}, \label{K_E}%
\end{equation}
we have%
\begin{subequations}
\begin{align}
c_{L}^{2}  &  =\frac{E}{\rho(1-\sigma^{2})},\\
c_{T}^{2}  &  =\frac{E}{2\rho(1+\sigma)}.
\end{align}
Rather than using the Poisson ratio directly it is convenient to introduce the
parameter for the ratio of wave speeds%
\end{subequations}
\begin{equation}
r=\frac{c_{L}}{c_{T}}=\sqrt{\frac{2}{1-\sigma}}>1. \label{wave_ratio}%
\end{equation}
In terms of the component displacements $u$, $v$ and the parameter $r$ the
elastic wave equation can be written
\begin{subequations}
\label{inplane_waves}%
\begin{align}
\frac{1}{c_{T}^{2}}\frac{\partial^{2}u}{\partial t^{2}}  &  =r^{2}%
\frac{\partial^{2}u}{\partial x^{2}}+\frac{\partial^{2}u}{\partial y^{2}%
}+(r^{2}-1)\frac{\partial^{2}v}{\partial x\partial y},\\
\frac{1}{c_{T}^{2}}\frac{\partial^{2}v}{\partial t^{2}}  &  =\frac{\partial
^{2}v}{\partial x^{2}}+r^{2}\frac{\partial^{2}v}{\partial y^{2}}%
+(r^{2}-1)\frac{\partial^{2}u}{\partial x\partial y}.
\end{align}

A result that we will find useful later is for the component displacements in
the solutions for the propagating waves
\end{subequations}
\begin{equation}
\mathbf{u}=\left\{
\begin{tabular}
[c]{ll}%
$u_{0}(1,-k_{Tx}/k_{Ty})e^{i(\mathbf{k}_{T}\cdot\mathbf{x}-\omega t)}$ &
transverse wave\\
$u_{0}(1,k_{Ly}/k_{Lx})e^{i(\mathbf{k}_{L}\cdot\mathbf{x}-\omega t)}$ &
longitudinal wave
\end{tabular}
\ \right.  \text{.} \label{inplane_components}%
\end{equation}

We note for completeness that in a \emph{three} dimensional
elastic medium, waves polarized in the $xy$ plane having no $z$
dependence are also described by an effective 2-dimensional
elasticity theory as in Eq.~(\ref{stress-strain2}) but now with
effective elastic constants $\bar
{K}=K+\frac{1}{3}\mu$, $\bar{\mu}=\mu$ so that $c_{L}^{2}/c_{T}^{2}%
=2(1-\sigma))/(1-2\sigma)$. The difference in the effective elastic constants
in the two cases arises because in the three dimensional medium restricted to
no $z$-dependence, there can be no expansion in the $z$-direction to relieve
the stresses set up by the strain in the $xy$ plane. In many elasticity
textbooks, the ``two-dimensional elasticity'' and ``thin-plate theory''
discussed correspond to this case. We can relate these results to the thin
plate geometry we are considering by appropriately transforming the effective
elastic constants.

For a finite plate we must apply stress free boundary conditions at the side
edges%
\begin{equation}
\hat{n}\cdot\mathbf{T}^{(2)}=0
\end{equation}
with $\hat{n}$ the normal to the edge. For waves propagating in the $x$
direction in a long finite plate of width $w$ the no-stress boundary condition
at the edges of the plate at $y=\pm w/2$ are therefore
\begin{subequations}
\label{Eq_ElasticBC}%
\begin{align}
T_{xy}^{(2)}  &  =0=\bar{\mu}\left(  \frac{\partial u}{\partial y}%
+\frac{\partial v}{\partial x}\right)  ,\\
T_{yy}^{(2)}  &  =0=\bar{\mu}\left(  r^{2}\frac{\partial v}{\partial y}%
+(r^{2}-2)\frac{\partial u}{\partial x}\right)  .
\end{align}

The boundary conditions have the effect of reflecting incident longitudinal
waves into both longitudinal and transverse reflected waves, so that these
waves become coupled in the finite geometry, leading to a complicated
dispersion relationship. The solutions propagating in the $x$ direction
decouple into either even or odd signature with respect to $y$ reflection, and
take the form (using Eqs.~(\ref{inplane_components}))%
\end{subequations}
\begin{subequations}
\begin{align}
u^{(e)}  &  =\left[  a_{T}^{(e)}\cos(\chi_{T}y)+a_{L}^{(e)}\cos(\chi
_{L}y)\right]  e^{i(kx-\omega t)}\\
v^{(e)}  &  =\left[  (-ik/\chi_{T})a_{T}^{(e)}\sin(\chi_{T}y)+(i\chi
_{L}/k)a_{L}^{(e)}\sin(\chi_{L}y)\right]  e^{i(kx-\omega t)}%
\end{align}
and%
\end{subequations}
\begin{subequations}
\begin{align}
u^{(o)}  &  =\left[  a_{T}^{(o)}\sin(\chi_{T}y)+a_{L}^{(o)}\sin(\chi
_{L}y)\right]  e^{i(kx-\omega t)}\\
v^{(o)}  &  =\left[  (ik/\chi_{T})a_{T}^{(o)}\cos(\chi_{T}y)-(i\chi
_{L}/k)a_{L}^{(o)}\cos(\chi_{L}y)\right]  e^{i(kx-\omega t)}%
\end{align}
where $\chi_{T,L}$ are given through the dispersion relation%
\end{subequations}
\begin{equation}
K^{2}=\omega^{2}/c_{T}^{2}=k^{2}+\chi_{T}^{2}=r^{2}(k^{2}+\chi_{L}^{2})
\end{equation}
and we have defined $K=\omega/c_{T}$, the wave number of a \emph{transverse
wave} at the frequency $\omega$ in an infinite plate. The values of
$\chi_{T,L}$ may be real or imaginary. Note that each wave combines both shear
($\chi_{T}$) and compressional ($\chi_{L}$) distortions, which are mixed by
the reflection of the plane waves off the edges.

The amplitudes $a_{T,L}^{(e),(o)}$ must be adjusted to satisfy the
boundary conditions at $y=\pm w/2$, Eqs.~(\ref{Eq_ElasticBC}).
Substituting into these
conditions leads to a system of two homogeneous equations for $a_{T}%
^{(e)},a_{L}^{(e)}$ (and two for $a_{T}^{(o)},a_{L}^{(o)}$) and so a
consistency condition that leads to a transcendental equation for $\omega$ for
each $k,$ known as the Rayleigh-Lamb equations\footnote{The Rayleigh-Lamb
equations usually occur in the somewhat different context of the analysis of
the modes in a plate which is considered infinite in the $xy$ plane, with
propagation in the $x$ direction and no dependence on the $y$ coordinate. The
wave numbers $\chi_{T,L}$ then give the variation across the thickness of the
plate. The equations take the same form, with the wave speed ratio $r$ given
by the expression for two dimensional elasticity theory, $r=\sqrt
{2(1-\sigma))/(1-2\sigma)}$. For example Rego and Kirczenow \cite{Rego98} plot
dispersion curves for these modes across the thickness of an infinite plate.}.%

\begin{figure}
[tbh]
\begin{center}
\includegraphics[
width=3.9972in
]%
{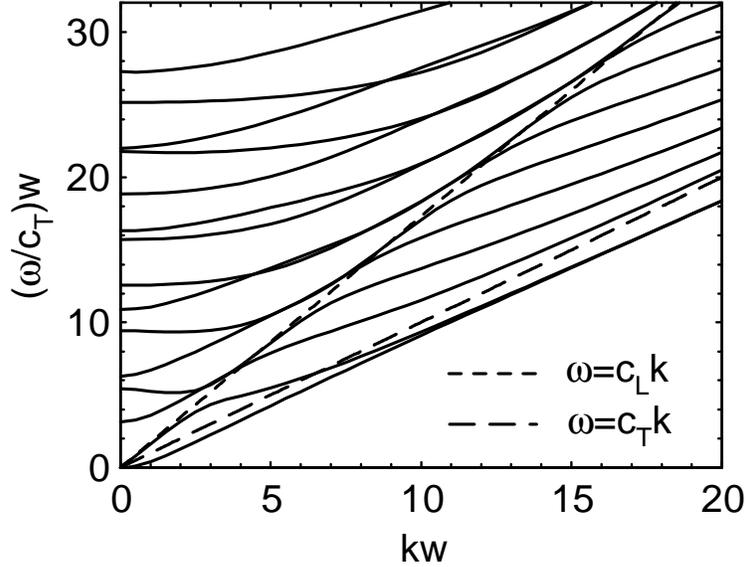}%
\caption{Dispersion relation of in-plane polarized modes for the thin plate.
The value of the Poisson ratio is $\sigma=0.33$.}%
\label{Fig_inplane_1}%
\end{center}
\end{figure}
\begin{figure}
[tbh]
\begin{center}
\includegraphics[
height=3.4938in,
width=3.9972in
]%
{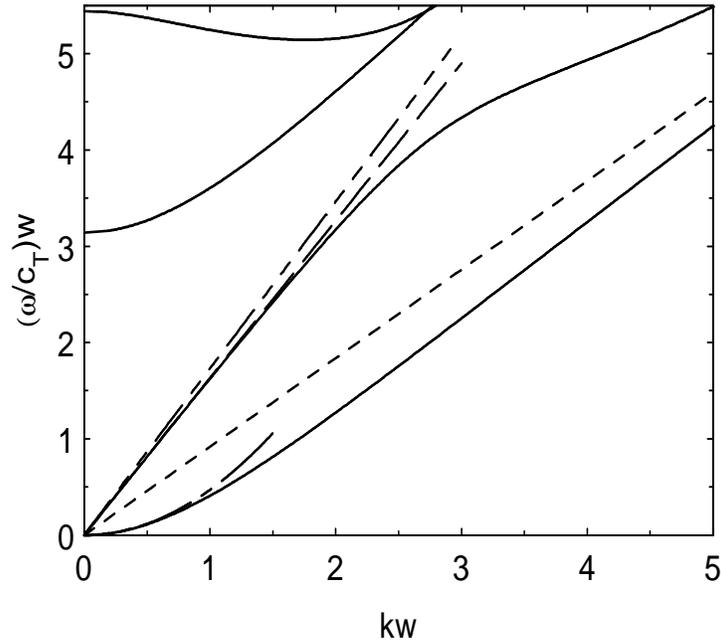}%
\caption{Dispersion curves for the long wavelength inplane modes of a thin
plate beam of width $w$ for Poisson ratio $1/3$ (wave speed ratio $\sqrt{3}$).
Solid lines: numerical results; long dashed line: small $k$ linear dispersion
for compression mode; long-short-short dashed: linear dispersion for
longitudinal wave in infinite plate; short dashed: linear dispersion for edge
mode; long-short dashed: quadratic dispersion for small $k$ bending mode. Note
the anomalous dispersion of the fourth mode at small $k$.}%
\label{Fig_inplane_2}%
\end{center}
\end{figure}

For the even signature modes the transcendental equation is
\begin{equation}
(k^{2}-\chi_{T}^{2})^{2}\tan\frac{\chi_{T}w}{2}+4k^{2}\chi_{T}\chi_{L}%
\tan\frac{\chi_{L}w}{2}=0. \label{transcendental_even}%
\end{equation}
For the odd signature modes the transcendental equation is
\begin{equation}
4k^{2}\chi_{T}\chi_{L}\tan\frac{\chi_{T}w}{2}+(k^{2}-\chi_{T}^{2})^{2}%
\tan\frac{\chi_{L}w}{2}=0. \label{transcendental_odd}%
\end{equation}
For a given value of the speed ratio $r$ these equations can be
solved numerically for $\omega(k)$. The spectrum for the value of
$r=\sqrt{3}$ corresponding to GaAs ($\sigma\simeq1/3$) is shown in
Fig.~(\ref{Fig_inplane_2}).

At large values of $k$ the slopes of the curves (except the lowest two)
asymptote to $c_{L}$ or $c_{T}$ corresponding to the freely propagating waves
in the plate. The lowest two modes asymptote to a slope $c_{S}$ less than both
$c_{T}$ and $c_{L}$. In this case for large $k$ we have $\tan(\chi
_{T,L}w/2)\rightarrow i$ in Eqs.~(\ref{transcendental_even}%
,\ref{transcendental_odd}) so that the slope $c_{S}=r_{S}c_{T}$ is given by
the solution of%
\begin{equation}
4\sqrt{1-r_{S}^{2}}\sqrt{1-r_{S}^{2}/r^{2}}=(2-r_{S}^{2})^{2}.
\label{Eq_Rayleigh}%
\end{equation}
This is an edge wave analogous to the Rayleigh wave on the surface of a three
dimensional slab of material. For $r=\sqrt{3}$ Eq.~(\ref{Eq_Rayleigh}) gives
$c_{S}=\sqrt{2-2/\sqrt{3}}c_{T}.$

The values of the finite-frequency intercepts of the ``waveguide''
modes for $k\rightarrow0$ can also be calculated analytically. For
the even mode intercepts Eq.~(\ref{transcendental_even}) is
satisfied at $k\rightarrow0$ by $\tan(\chi_{T}w/2)=0$ or
$\tan(\chi_{L}w/2)\rightarrow\infty$. Similarly
Eq.~(\ref{transcendental_odd}) is satisfied by
$\tan(\chi_{L}w/2)=0$ or $\tan (\chi_{T}w/2)\rightarrow\infty$.
Thus the zero wave number intercepts are given by the simple
expressions for transverse and longitudinal wave propagation
\begin{equation}
\omega_{n}^{(T)}=n\pi c_{T}/w,\quad\omega_{n}^{(L)}=rn\pi c_{T}/w=n\pi
c_{L}/w.
\end{equation}
(These simple results hold because for $k=0$ there is no interconversion of
longitudinal and transverse waves on reflection at the edges.) The shapes of
the curves between $k=0$ and the large $k$ asymptotes are quite complicated,
with various mode crossings and regions of anomalous dispersion $d\omega/dk<0$
for example.%

We are particularly interested in the long wavelength modes
$k\rightarrow0$. The dispersion relation in this limit can be
found by Taylor expansion of the $\tan$ functions in
Eqs.~(\ref{transcendental_even},\ref{transcendental_odd}). The
even mode tends to
\begin{equation}
\omega\rightarrow2\sqrt{1-r^{-2}}c_{T}k=\sqrt{E/\rho}\,k.
\end{equation}
This agrees with the usual expression for the stretching mode of a rod. This
is not the same as the dispersion for the bulk longitudinal modes in the thin
plate, $\omega=c_{L}k$, but the speeds are quite close for $\sigma=0.33$. On
the other hand the odd mode gives a \emph{quadratic} dispersion,
characteristic of bending modes of beams,%
\begin{equation}
\omega\rightarrow\sqrt{\frac{r^{2}-1}{3r^{2}}}wc_{T}k^{2}.
\end{equation}
(This is given by expanding the $\tan$ functions up to cubic order.) Rod
bending theory gives the expression $\omega/k^{2}=\sqrt{EI/\rho A}$ with $I$
the areal moment of inertia about the midline, and $A$ the cross section area.
For the rectangular beam we have $I/A=w^{2}/12$, and using $(1-r^{-2}%
)c_{T}^{2}=E/2\rho$ shows the correspondence.

\subsubsection{Flexural modes}

\label{Subsec_Flexural}

The flexural modes are most easily derived by using relationships
such as Eq.~(\ref{eliminate_uz}) to derive an expression for the
energy of transverse
displacements \cite{Landau86}%
\begin{equation}
F=\frac{1}{2}D\int\int\left[  \left(  \nabla_{\perp}^{2}w\right)
^{2}+2(1-\sigma)\left\{  \left(  \frac{\partial^{2}w}{\partial x\partial
y}\right)  ^{2}-\frac{\partial^{2}w}{\partial x^{2}}\frac{\partial^{2}%
w}{\partial y^{2}}\right\}  \right]  dxdy
\end{equation}
where%
\begin{equation}
D=\frac{Ed^{3}}{12(1-\sigma^{2})}%
\end{equation}
is the flexural rigidity of the plate of thickness $d$. The equation of motion
and boundary conditions are given by taking the variation of the energy with
respect to displacements $w(x,y)$. For a region with rectangular boundaries at
$x=\pm a/2$ and $y=\pm b/2$ the variation is%
\begin{multline}
\delta F=D\int\int\nabla_{\perp}^{4}\omega\,\delta w\,dxdy-\\
\left.  \int dy\left\{  \delta w\left[  \frac{\partial^{3}w}{\partial x^{3}%
}+(2-\sigma)\frac{\partial^{3}w}{\partial x\partial y^{2}}\right]
-\frac{\partial\delta w}{\partial x}\left[  \frac{\partial^{2}w}{\partial
x^{2}}+\sigma\frac{\partial^{2}w}{\partial y^{2}}\right]  \right\}  \right|
_{x=-a/2}^{x=a/2}-\\
\left.  \int dx\left\{  \delta w\left[  \frac{\partial^{3}w}{\partial y^{3}%
}+(2-\sigma)\frac{\partial^{3}w}{\partial x^{2}\partial y}\right]
-\frac{\partial\delta w}{\partial y}\left[  \frac{\partial^{2}w}{\partial
y^{2}}+\sigma\frac{\partial^{2}w}{\partial x^{2}}\right]  \right\}  \right|
_{y=-b/2}^{y=b/2}.
\end{multline}
The first term gives the effective force per unit area on the plate, and hence
the equation of motion%
\begin{equation}
\rho d\frac{\partial^{2}w}{\partial t^{2}}+D\nabla_{\perp}^{4}w=0.
\end{equation}
The last two terms on the right hand side are boundary terms given by
integration by parts. Physically they give the work done at the boundaries by
the vertical force per length of boundary $V_{i}$ against the vertical
displacement $\delta w$ and by the moment per unit length $M_{i}$ against the
angular displacement of the plate\footnote{The sign convention for the moments
$M_{x},M_{y}$ is that $M_{i}$ is positive if it tends to produce compression
in the \emph{negative} $z$ side of the plate. The angular displacements
$\theta_{x},\theta_{y}$ are defined with the same convention. This is the
usual definition in the elasticity literature \cite{Timoshenko61}.}
$-\vec{\nabla}_{i}w$. Thus at $x=a/2$ we have%
\begin{subequations}
\begin{align}
V_{x}  &  =-D\frac{\partial}{\partial x}\left(  \frac{\partial^{2}w}{\partial
x^{2}}+(2-\sigma)\frac{\partial^{2}w}{\partial y^{2}}\right)  ,\\
M_{x}  &  =-D\left(  \frac{\partial^{2}w}{\partial x^{2}}+\sigma
\frac{\partial^{2}w}{\partial y^{2}}\right)  ,
\end{align}
and at $y=b/2$
\end{subequations}
\begin{subequations}
\label{flex_bc_y}%
\begin{align}
V_{y}  &  =-D\frac{\partial}{\partial y}\left(  \frac{\partial^{2}w}{\partial
y^{2}}+(2-\sigma)\frac{\partial^{2}w}{\partial x^{2}}\right)  ,\\
M_{y}  &  =-D\left(  \frac{\partial^{2}w}{\partial y^{2}}+\sigma
\frac{\partial^{2}w}{\partial x^{2}}\right)  .
\end{align}
For free edges, these quantities must be set to zero. In addition to the force
per unit length there are also point forces localized at the corners e.g. at
$x=a/2,y=\pm b/2$ \cite{Timoshenko61}%
\end{subequations}
\begin{equation}
F_{c}=\pm2D(1-\sigma)\frac{\partial^{2}w}{\partial x\partial y}.
\label{corner_force}%
\end{equation}
These must be included when we are calculating the total force acting across
the width of the beam, for example:%
\begin{align}
F  &  =\int_{-w/2}^{w/2}V_{x}\,dy+F_{c}(w/2)+F_{c}(-w/2)\\
&  =\frac{\partial}{\partial x}\int_{-w/2}^{w/2}M_{x}\,dy
\end{align}
and the latter equality shows the consistency with the macroscopic equation
for the rotational equilibrium of the beam (see section \ref{Sec_bending} below).

We now calculate the modes propagating in the $x$ direction $w^{(e)}%
,w^{(o)}\propto e^{i(kx-\omega t)}$ in the bridge of width $w$. Again the
modes have either even or odd signature with respect to $y$ reflections. Since
the wave equation is fourth order in the spatial derivatives, for each
frequency $\omega$ there are \emph{two} even or odd components. The solutions
to the wave equation are (even)%
\begin{equation}
w^{(e)}=\left[  a_{+}^{(e)}\cosh(\chi_{+}y)+a_{-}^{(e)}\cosh(\chi
_{-}y)\right]  e^{i(kx-\omega t)} \label{Eq_FlexEven}%
\end{equation}
and (odd)%
\begin{equation}
w^{(o)}=\left[  a_{+}^{(o)}\sinh(\chi_{+}y)+a_{-}^{(o)}\sinh(\chi
_{-}y)\right]  e^{i(kx-\omega t)} \label{Eq_FlexOdd}%
\end{equation}
where%
\begin{equation}
\chi_{\pm}=\sqrt{k^{2}\pm K^{2}},
\end{equation}
and we have written%
\begin{equation}
\sqrt{\rho d/D}\,\omega=K^{2}. \label{Omega_Ksq}%
\end{equation}%

\begin{figure}
[tbh]
\begin{center}
\includegraphics[
width=3.9946in
]%
{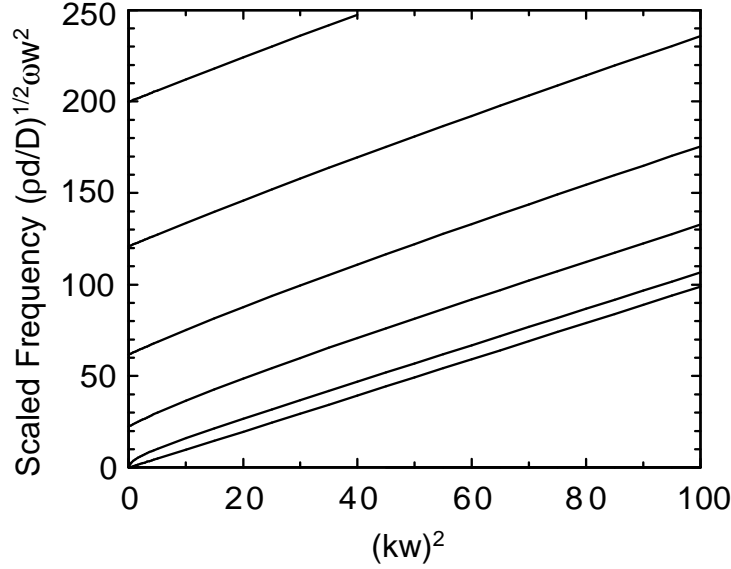}%
\caption{Dispersion relation of the flexural modes of a thin plate with
Poisson ratio $\sigma=0.33$. Note that the modes have approximately a
quadratic wave number dependence $\omega\simeq\omega_{c}^{2}+\alpha k^{2}$
with $\alpha\simeq\sqrt{D/\rho d}$, corresponding to the ``bulk'' flexural
wave. The two lower modes have a different form: one mode (the torsion mode)
has a linear dispersion relation at low frequencies $\omega\propto k$ and both
modes asymptote to $\omega=\beta k^{2}$ at large $k$ with $\beta<\alpha$
corresponding to an edge wave. It turns out that $\alpha\simeq\beta$ for the
particular value of $\sigma$ used here.}%
\label{Fig_flex_1}%
\end{center}
\end{figure}
Again the dispersion $\omega(k)$ and the ratio of amplitudes
$a_{+}/a_{-}$ are determined by the requirement of consistency
with the boundary conditions at
the edges $y=\pm w/2$ Eq.~(\ref{flex_bc_y}). This gives for the even modes%
\begin{equation}
\lbrack K^{2}+(1-\sigma)k^{2}]^{2}\chi_{-}\tanh(\chi_{-}w/2)=[K^{2}%
-(1-\sigma)k^{2}]^{2}\chi_{+}\tanh(\chi_{+}w/2) \label{flex_mode_even}%
\end{equation}
and for the odd modes%
\begin{equation}
\lbrack K^{2}+(1-\sigma)k^{2}]^{2}\chi_{-}\coth(\chi_{-}w/2)=[K^{2}%
-(1-\sigma)k^{2}]^{2}\chi_{+}\coth(\chi_{+}w/2). \label{flex_mode_odd}%
\end{equation}

For $\omega\rightarrow0$ we can expand the hyperbolic functions and solve
algebraic equations to determine the dispersion curve.%

\begin{figure}
[tbh]
\begin{center}
\includegraphics[
height=2.9776in,
width=3.9972in
]%
{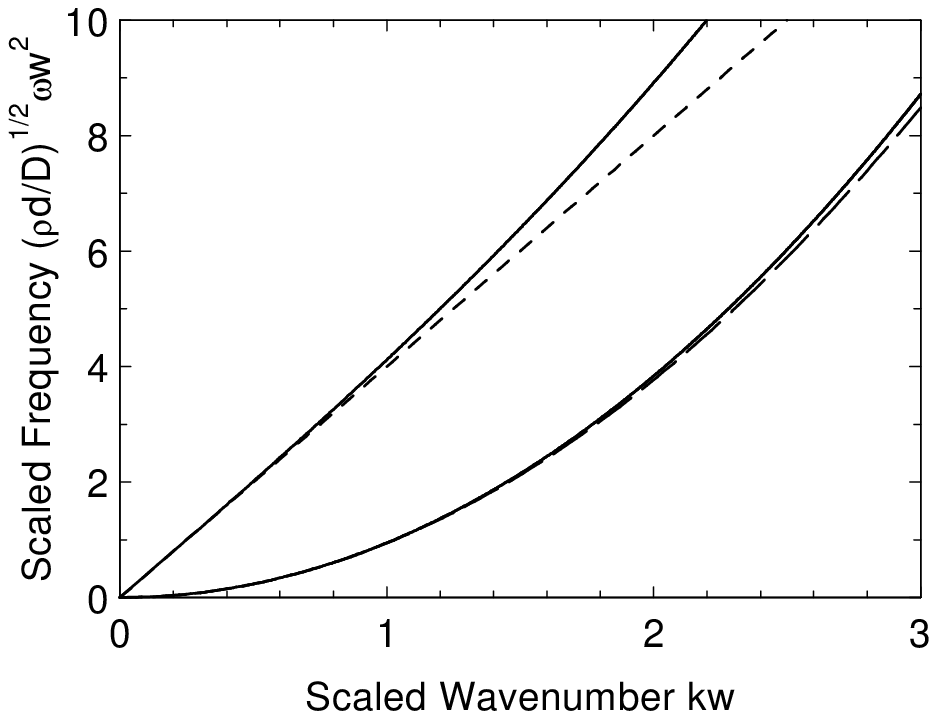}%
\caption{Dispersion relation as in Fig.~(\ref{Fig_flex_1}) but at small
$\omega,k$. Note that the abscissa is proportional to $k$ in this plot. Solid
lines: numerical results; long-dashed line: expected quadratic dispersion for
rod-bending mode; short-dashed line: expected linear dispersion for
rod-torsion mode.}%
\label{Fig_flex_2}%
\end{center}
\end{figure}
For the even mode this gives $K^{2}/k^{2}=\sqrt{1-\sigma^{2}}$
yielding the quadratic dispersion of the beam bending mode
\begin{equation}
\omega=\sqrt{\frac{D(1-\sigma^{2})}{\rho d}}k^{2}=\sqrt{\frac{E}{12\rho}%
}dk^{2},
\end{equation}
agreeing with the expression from simple rod theory. We can follow this mode
to large $k$ where we find again a quadratic dispersion but with a different
slope%
\begin{equation}
K^{2}/k^{2}\rightarrow\sqrt{\frac{1-3\sigma+2\sqrt{1-2\sigma(1-\sigma)}%
}{(1-\sigma)^{2}(3+\sigma)}}. \label{flex_Rayleigh}%
\end{equation}
This is again an edge wave (now an edge bending wave). The
intersections of the higher modes with the frequency axis are
given by $k^{2}\rightarrow0$, $\chi_{-}\rightarrow iK$,
$\chi_{+}\rightarrow K$ so that
Eq.~(\ref{flex_mode_even}) reduces to%
\begin{equation}
-\tan(Kw/2)=\tanh(Kw/2).
\end{equation}
The solutions are well approximated by $Kw=(3/2+2n)\pi$ (i.e. $\omega
=\sqrt{D/\rho d}(3/2+2n)^{2}\pi^{2}/w^{2}$) for $n=0,1\ldots$ (with an error
of less than $1/2\%$ for the worst case $n=0$).

For the odd mode Eq.~(\ref{flex_mode_odd}) gives for $\omega,k\rightarrow0$
the dispersion relation for the torsion mode
\begin{equation}
\omega=\sqrt{\frac{D}{\rho d}}2\sqrt{6(1-\sigma)}\frac{k}{w}=2\sqrt{\frac{\mu
}{\rho}}\frac{d}{w}k, \label{Eq_FlexTorsion}%
\end{equation}
which agrees with the usual result calculated in elastic rod theory. The large
$k$ asymptote of this mode is the same as Eq.~(\ref{flex_Rayleigh}). The
intersections of the higher odd modes with the frequency axis are from
Eq.~(\ref{flex_mode_odd}) given by%
\begin{equation}
\tan(Kw/2)=\tanh(Kw/2).
\end{equation}
The solutions are well approximated by $Kw=(5/2+2n)\pi$ (i.e. $\omega
=\sqrt{D/\rho d}(5/2+2n)^{2}\pi^{2}$) for $n=0,1\ldots$.

Combining the even and odd modes, the zero wave number frequency intercepts
can be written%
\begin{equation}
\omega_{n}\simeq c_{T}\frac{1}{\sqrt{6(1-\sigma)}}(\frac{3}{2}+n)^{2}\pi
^{2}d/w^{2}.
\end{equation}

The dispersion curves for $\sigma=1/3$ corresponding to GaAs are
shown in Figs.~(\ref{Fig_flex_1},\ref{Fig_flex_2}).

\subsection{Transmission coefficient in the infinite wavelength limit}

The transmission coefficients for the acoustic modes in the long wavelength
limit and for finite cavity width can be readily calculated by the wave
matching methods as for the scalar waves. We investigate the transmission of
very long wavelength modes at the abrupt junction between the bridge ($x<0$)
of width $w$ and the cavity ($x>0$) of finite width $W$. It is easiest to
evaluate the wave fields by simple macroscopic arguments. As well as providing
the long wavelength limit of the transmission coefficients, these results also
provide the basis for calculating the leading order finite $k$ corrections,
following the same methods as in Sec. \ref{secMPscalar}.

\subsubsection{Compression modes}

The compressional (extension) modes for $k\rightarrow0$ are given by the
simple one dimensional calculation:%
\begin{equation}
\frac{\partial^{2}u}{\partial t^{2}}=c_{E}^{2}\frac{\partial^{2}u}{\partial
x^{2}}.
\end{equation}
where $c_{E}^{2}=E/\rho$.

For $x<0$ we have incident and reflected waves%
\begin{equation}
u=(e^{ikx}+re^{-ikx})e^{-i\omega t}%
\end{equation}
and for $x>0$ a single transmitted wave%
\begin{equation}
u=te^{ikx}e^{-i\omega t}%
\end{equation}
where $\omega/k=c_{E}$. We match the displacement $u$ and the total force
either side of the interface%
\begin{align}
1+r  &  =t,\\
ikw(1-r)  &  =ikWt.
\end{align}
Note that the force matching requires that the end surface of the cavity be
stress free. This gives%
\begin{equation}
t=\frac{1}{1+h},\quad r=-\frac{h-1}{h+1}.
\end{equation}
where $h$ is the width ratio $h=W/w$. In the limit $h\rightarrow\infty$ we
find $r\rightarrow-1$, $t\rightarrow0$, i.e. perfect reflection with a sign
change of the displacement. This implies that at the junction $x=0$ we have
$u\simeq0$ and the stress $E\partial u/\partial x\simeq2ikEe^{-i\omega t}$.

\subsubsection{Bending modes}

\label{Sec_bending}

At long wavelengths the bending modes are given by equations for the total
force $F$ and the total moment $M$ on each cross-section (see
\cite{Timoshenko61}). The moment from opposite forces on each end of a small
element of the beam must cancel the net moment from the forces on the faces%
\begin{equation}
F=\frac{\partial M}{\partial x}.
\end{equation}
(Since the moments scale as the length of the element $\delta x$, whereas the
moment of inertia scales as $\delta x^{2}$, there is no inertial term in this
equation). The net force on an element gives its acceleration%
\begin{equation}
\rho A\frac{\partial^{2}u}{\partial t^{2}}=\frac{\partial F}{\partial x},
\end{equation}
where $u$ is the bending displacement and $A$ is the cross section area. The
moment $M$ is given by the moment of the tensile stresses due to the extension
and compression of the beam from its curvature%
\begin{equation}
M=-EI\frac{\partial^{2}u}{\partial x^{2}},
\end{equation}
where $I$ is the areal moment of inertia of the beam around the midline normal
to the displacement. These equations together give the equation of motion%
\begin{equation}
\frac{\partial^{2}u}{\partial t^{2}}+\frac{EI}{\rho A}\frac{\partial^{4}%
u}{\partial x^{4}}=0.
\end{equation}
The dispersion relation is quadratic, $\omega=\alpha k^{2}$ with $\alpha
=\sqrt{EI/\rho A}$. For a rectangular beam $I/A=d^{2}/12$ with $d$ the
thickness of the beam in the direction of the displacement. As we saw in the
previous section, this reproduces the long wavelength limit of the acoustic
bending modes calculated in thin plate theory.

At a frequency $\omega$, as well as the propagating modes at wave number
$\pm\sqrt{\omega/\alpha}$ there are also evanescent modes with decay rate
$\pm\sqrt{\omega/\alpha}$: the modes localized at the junction and decaying to
$\pm\infty$ must be included in the mode transmission problem.

For the bending mode with displacement normal to the plane the dispersion
relation is the same in the bridge and cavity. Thus for an incident wave
$e^{i(kx-\omega t)}$ in the bridge with $k=\sqrt{\omega/\alpha}$ we have%
\begin{equation}
u=\left\{
\begin{tabular}
[c]{ll}%
$e^{ikx}+re^{-ikx}+ae^{kx}$ & $x<0$\\
$te^{ikx}+be^{-kx}$ & $x>0$%
\end{tabular}
\ \ \right.  .
\end{equation}
At the junction we require continuity of the displacement $u$, the rotation
angle $\partial u/\partial x$, the total moment $-EI\partial^{2}u/\partial
x^{2}$ and the total force $-EI\partial^{3}u/\partial x^{3}$. This gives the
matrix equation%
\begin{equation}
\left[
\begin{array}
[c]{cccc}%
1 & -1 & 1 & -1\\
1 & 1 & i & i\\
1 & -h & -1 & h\\
1 & h & -i & -ih
\end{array}
\right]  \left[
\begin{array}
[c]{c}%
r\\
t\\
a\\
b
\end{array}
\right]  =\left[
\begin{array}
[c]{c}%
-1\\
1\\
-1\\
1
\end{array}
\right]  ,
\end{equation}
where $h=I_{c}/I_{b}=W/w$ is the ratio of the appropriate moment of inertia
for the cavity and bridge. In the limit of large $h$ the solution is easily
found to be
\begin{subequations}
\label{flex_bend_junction_zeroth}%
\begin{align}
r  &  =i,\\
t  &  =4/h,\\
a  &  =-(1+i),\\
b  &  =2(1-i)/h.
\end{align}

For the bending mode with displacements in the plane of the plate the
dispersion relation is different in bridge and cavity. For frequency $\omega$
the wave numbers in the bridge and cavity are $k_{b},k_{c}$ with%
\end{subequations}
\begin{equation}
k_{c}/k_{b}=(\alpha_{b}/\alpha_{c})^{1/2}=(w/W)^{1/2}.
\end{equation}
On the other hand the ratio of the moments of inertia is%
\begin{equation}
I_{c}/I_{b}=(W/w)^{3}\text{.}%
\end{equation}
Thus the continuity of the displacement $u$, the rotation angle $\partial
u/\partial x$, the total moment $-EI\partial^{2}u/\partial x^{2}$ and the
total force $-EI\partial^{3}u/\partial x^{3}$. This gives the matrix equation
(writing $\bar{h}=\sqrt{W/w}$)%
\begin{equation}
\left[
\begin{array}
[c]{cccc}%
1 & -1 & 1 & -1\\
1 & \bar{h}^{-1} & i & i\bar{h}^{-1}\\
1 & -\bar{h}^{4} & -1 & \bar{h}^{4}\\
1 & \bar{h}^{3} & -i & -i\bar{h}^{3}%
\end{array}
\right]  \left[
\begin{array}
[c]{c}%
r\\
t\\
a\\
b
\end{array}
\right]  =\left[
\begin{array}
[c]{c}%
-1\\
1\\
-1\\
1
\end{array}
\right]  ,
\end{equation}
In the limit of large $\bar{h}$ the solution is%
\begin{subequations}
\begin{align}
r  &  =i,\\
t  &  =2/\bar{h}^{3},\\
a  &  =-(1+i),\\
b  &  =2/\bar{h}^{3}.
\end{align}

With these expression in both cases we have at $x=0$ to leading order in
$h^{-1}$ or $\bar{h}^{-1}$
\end{subequations}
\begin{subequations}
\label{bend_junction_zeroth}%
\begin{align}
u  &  =0,\\
\partial u/\partial x  &  =0,\\
\partial^{2}u/\partial x^{2}  &  =-2(1+i),\\
\partial^{3}u/\partial x^{3}  &  =-2(1+i),
\end{align}
so that the displacements, $u$ and the angle $\partial u/\partial x$, tend to
zero, but the corresponding stresses are large. Note that although the force
and moment are out of phase for the single wave $e^{i(kx-\omega t)}$, because
of the evanescent waves near the junction they are in phase at the junction
plane $x=0$. Equations (\ref{bend_junction_zeroth}) become the zeroth order
input for the radiation calculation at non-zero wave number.

\subsubsection{Torsion modes}

The long wavelength limit of torsion waves is described by the one dimensional
wave equation giving the angular acceleration in terms of the torque%
\end{subequations}
\begin{equation}
I\frac{\partial^{2}\theta}{\partial t^{2}}=C\frac{\partial^{2}\theta}{\partial
x^{2}}.
\end{equation}
Here $C$ is the torsional rigidity giving the torque on a section due to the
twist $\tau=\partial\theta/\partial x$%
\begin{equation}
\text{torque}=C\tau.
\end{equation}
It is given by \cite{Landau86}%
\begin{equation}
C=4\mu\int\chi\,dxdy
\end{equation}
where $\chi$ satisfies the equation in the cross section%
\begin{equation}
\nabla_{\perp}^{2}\chi=-1
\end{equation}
and $\chi=0$ on the boundaries. The form of the solution $\chi$ is the same as
the profile of the flow of a viscous fluid through the section and $C$ is then
proportional to the integrated flux. For the thin plate geometry with
thickness $d$ and width $w\gg d$, the value of $C$ is $\frac{1}{3}\mu d^{3}w$
and value of $I$ is $\frac{1}{12}dw^{3}$, so that the ratio of propagation
speeds in bridge and cavity is again the width ratio $h=W/w$.

It is interesting to evaluate the stress distribution for the thin plate. The
stresses in the $(y,z)$ section are given in terms of $\chi$ by
\cite{Landau86}%
\begin{subequations}
\begin{align}
\sigma_{yx}  &  =2\mu\tau\partial\chi/\partial z\\
\sigma_{zx}  &  =-2\mu\tau\partial\chi/\partial y.
\end{align}
The solution for $\chi$ is analogous to Poiseuille flow, so that%
\end{subequations}
\begin{equation}
\chi=\frac{1}{8}(d^{2}-4z^{2})
\end{equation}
except within a distance $\delta\sim d$ from the side wall, where $\chi$ must
decrease to zero. Thus there is a distributed stress acting in the $y$
direction%
\begin{equation}
\sigma_{yx}\simeq-2\mu\tau z
\end{equation}
and a stress in the $z$ direction that is effectively localized (within a
distance $d)$ at the edge%
\begin{equation}
\sigma_{zx}\simeq\frac{1}{4}\mu\tau(d^{2}-4z^{2})\delta(y-w/2).
\end{equation}
This localized stress corresponds to the corner forces Eq.~(\ref{corner_force}%
) in the thin plate theory.

For an incident torsion wave $\theta=e^{i(k_{b}x-\omega t)}$ in the bridge for
$x<0$ we have incident and reflected waves%
\begin{equation}
\theta=(e^{ik_{b}x}+re^{-ik_{b}x})e^{-i\omega t}%
\end{equation}
and for $x>0$ a single transmitted wave%
\begin{equation}
\theta=te^{ik_{c}x}e^{-i\omega t}%
\end{equation}
where $k_{c}/k_{b}=W/w$. The matching of the angular displacement $\theta$ and
torque $C\partial\theta/\partial x$ for an incident wave gives%
\begin{subequations}
\begin{align}
1+r  &  =t\\
ik_{b}w(1-r)  &  =ik_{c}Wt
\end{align}
so that%
\end{subequations}
\begin{equation}
t=\frac{1}{1+h^{2}},\quad r=-\frac{h^{2}-1}{h^{2}+1}.
\end{equation}
with $h=W/w$. In the limit $h\rightarrow\infty$ we find $r\rightarrow-1$,
$t\rightarrow0$, implying $\theta(x=0)\simeq0$ and the stress $C\partial
\theta/\partial x\simeq2ikCe^{-i\omega t}$.

\subsection{Transmission coefficient for small k}

In this section we calculate the small wave vector asymptotic limit of the
transmission coefficient from the four acoustic modes of the beam into the
cavity. The method follows that of section \ref{secMPscalar}. Thus we
calculate the radiation from oscillating stresses $s(y)e^{-i\omega t}$ on the
edge of the cavity. The stresses are calculated as the stresses arising on the
ends of the bridge for zero displacement boundary conditions, as follows from
the analyses in the previous section of the modes at infinite wavelengths
coupling into a cavity of finite width. For the long wavelength value of the
transmission coefficients, only the radiation by the integrated stress $\int
s(y)dy$ for the even parity modes, or the integrated first moment $\int
s(y)ydy$ for the odd parity modes, is needed. The Lamb problem of the
radiation from surface sources into an elastic half space has been much
studied in the literature, for example see \cite{Miller54}, and the reader is
referred there for a more exhaustive discussion of this aspect of the
calculation. The details of the calculations are quite complicated, and the
reader may choose to skip these sections and refer to the discussion of the
results tabulated in section \ref{Sec_Applications} and the summary in table
\ref{Table1} there.

\subsubsection{In-plane Compression}

For a compressional wave in the bridge of unit incident amplitude in the
displacement, the oscillating end of the bridge acts as a stress source on the
cavity face of amplitude $2iEk$ over the source region $|y|<w/2$, embedded in
the otherwise stress free line $x=0$. The solutions to the wave equations
(\ref{inplane_waves}) in the cavity can be written (cf.
Eq.~(\ref{inplane_components}))%
\begin{subequations}
\begin{align}
u  &  =\frac{1}{2\pi}\int_{-\infty}^{\infty}[a_{T}(\zeta)e^{ik_{T}x}%
+a_{L}(\zeta)e^{ik_{L}x}]e^{i\zeta y}e^{-i\omega t}d\zeta\\
v  &  =\frac{1}{2\pi}\int_{-\infty}^{\infty}[-(k_{T}/\zeta)a_{T}%
(\zeta)e^{ik_{T}x}+(\zeta/k_{L})a_{L}(\zeta)e^{ik_{L}x}]e^{i\zeta
y}e^{-i\omega t}d\zeta
\end{align}
where $k_{T}$ and $k_{L}$ are the $x$-components of the wave vectors of the
transverse and longitudinal components%
\end{subequations}
\begin{equation}
k_{T}=\left\{
\begin{tabular}
[c]{ll}%
$\sqrt{K^{2}-\zeta^{2}}$ & $|\zeta|\leq K$\\
$i\sqrt{\zeta^{2}-K^{2}}$ & $|\zeta|>K$%
\end{tabular}
\ \ \ \ \ \right.  ,\quad k_{L}=\left\{
\begin{tabular}
[c]{ll}%
$\sqrt{K^{2}/r^{2}-\zeta^{2}}$ & $|\zeta|\leq K/r$\\
$i\sqrt{\zeta^{2}-K^{2}/r^{2}}$ & $|\zeta|>K/r$%
\end{tabular}
\ \ \ \ \ \right.
\end{equation}
with $K^{2}=\omega^{2}/c_{T}^{2}$. (The signs chosen correspond to waves
propagating away or exponentially decaying.) The amplitudes $a_{L,T}$ are
fixed by matching to the normal and tangential stress sources $\Sigma
_{n},\Sigma_{t}$ for $|y|<1/2$
\begin{subequations}
\label{in_plane_stress}%
\begin{align}
\bar{\mu}\left(  r^{2}\frac{\partial u}{\partial x}+(r^{2}-2)\frac{\partial
v}{\partial y}\right)   &  =\Sigma_{n},\\
\bar{\mu}\left(  \frac{\partial u}{\partial y}+\frac{\partial v}{\partial
x}\right)   &  =\Sigma_{t}.
\end{align}
In the present case $\Sigma_{n}=2iEke^{-i\omega t}$ for unit incident wave
amplitude and $\Sigma_{t}=0$. Both components of the stress are zero for
$|y|>w/2$. Taking the leading order expansion in $Kw$ for the Fourier
transform of the source stress as in the scalar calculation, section
\ref{secMPscalar}, gives for the Fourier components%
\end{subequations}
\begin{subequations}
\begin{align}
r^{2}(ik_{T}a_{T}+ik_{L}a_{L})+(r^{2}-2)(-ik_{T}a_{T}+i\zeta^{2}a_{L}/k_{L})
&  =\Sigma_{n}w/\mu\\
i\zeta(a_{T}+a_{L})+(-ik_{T}^{2}a_{T}/\zeta+i\zeta a_{L})  &  =0
\end{align}
These equations are readily solved for $a_{L,T}$ from which we can calculate
$u(x=0)$ for unit driving stress%
\end{subequations}
\begin{equation}
\frac{u(x=0)}{\Sigma_{n}}=-\frac{r^{2}w}{\bar{\mu}}\frac{i}{\pi}\int
_{0}^{\infty}\frac{\kappa_{L}(\xi)}{F_{o}(\xi)}d\xi
\end{equation}
where $\xi=r\zeta/K$ is the $y$-wave vector scaled by the \emph{longitudinal}
wave number $K/r=\omega/c_{L}$,%
\begin{equation}
F_{0}(\xi)=(2\xi^{2}-r^{2})^{2}+4\xi^{2}\kappa_{T}(\xi)\kappa_{L}(\xi)
\end{equation}
and%
\begin{equation}
\kappa_{T}=\left\{
\begin{tabular}
[c]{ll}%
$\sqrt{r^{2}-\xi^{2}}$ & $|\xi|\leq r$\\
$i\sqrt{\xi^{2}-r^{2}}$ & $|\xi|>r$%
\end{tabular}
\ \ \ \ \ \right.  ,\quad\kappa_{L}=\left\{
\begin{tabular}
[c]{ll}%
$\sqrt{1-\xi^{2}}$ & $|\xi|\leq1$\\
$i\sqrt{\xi^{2}-1}$ & $|\xi|>1$%
\end{tabular}
\ \ \ \ \ \right.
\end{equation}

(This result is analogous to Eq.~(123) of Miller and Pursey \cite{Miller54}
who calculate the average displacement at the aperture per unit oscillating
stress for a \emph{line} on a three dimensional half space. Indeed we can use
their result if we express it in terms of the ratio of wave speeds, and the
elastic constant $\mu$ which retains its significance unchanged between the
two geometries. The translation from their (MP) notation is then
$u_{x,MP}\rightarrow(\omega/c_{L})u_{x}$, $a_{MP}\rightarrow(\omega/c_{L}%
)w/2$, $\mu_{MP}\rightarrow r$, $c_{44,MP}\rightarrow\bar{\mu}$. Note
carefully that the usage of ``$\mu$'' is different in their work and ours. We
have also taken the leading order term in $\omega w/c_{L}$ by making the
replacement $e^{i\xi y}\rightarrow1$ for $|y|<w/2$.)

For unit incident wave in the beam the longitudinal stress at the aperture is
$\Sigma_{n}=2iEke^{-i\omega t}$ and the power radiated is $\frac{1}%
{2}w\operatorname{Re}(-\dot{u}\Sigma_{n}^{\ast})$. In the incident wave the
stress is $\Sigma_{w}=iEke^{-i\omega t}$ and the velocity $\dot{u}_{x}$ is
$-i\omega e^{-i\omega t}$ so that the incident power is $\frac{1}{2}wE\omega
k$. The transmission coefficient is therefore%
\begin{equation}
\mathcal{T}_{0}(\omega)=4(kw)\left[  \frac{4}{\pi}\frac{(1+\sigma)}%
{(1-\sigma)}\operatorname{Re}\int_{0}^{\infty}\frac{\kappa_{L}(\xi)}{F_{0}%
(\xi)}d\xi\right]
\end{equation}
where we have used $E/\bar{\mu}=2(1+\sigma)$ and $r^{2}=2/(1-\sigma)$. The
various contributions to the the integral are easily understood in terms of
the different waves radiated into the cavity. Remember that $\xi$ is the $y$
wave number of the radiated waves in units of $k_{L}$. There are contributions
to the integral for $0<\xi<r$ corresponding to the radiation of transverse and
longitudinal waves over all angles. In addition there is a contribution from
the residue of the pole at $F_{0}(\xi)=0$ which corresponds to the radiation
of \emph{edge} waves. Miraculously, the quantity in the square brackets
numerically evaluates to $1.0$ independent of $\sigma$ in the allowed range
$-1<\sigma<1/2$, so that%
\begin{equation}
\mathcal{T}_{0}(\omega)=4kw.
\end{equation}

\subsubsection{In-plane Bending}

For an incident bending wave with unit displacement amplitude $v$ in the
$y$-direction there are two sources of radiation into the cavity: the
oscillating moment $2\sqrt{2}EIk^{2}e^{-i\omega t}$ and the oscillating shear
(tangential) force $2\sqrt{2}EIk^{3}e^{-i\omega t}$ over the source width $w$
in the cavity wall. The moment can be described in terms of the normal stress
$2\sqrt{2}Ek^{2}e^{-i\omega t}y$ since $\int y^{2}dy=I$. The tangential force
has an additional power of $k\sim\sqrt{\omega}$ compared to this, but the
radiation efficiency of the normal force is reduced by a power of $k_{c}w$
with $k_{c}$ the wave number of a propagating mode in the cavity $k_{c}%
\sim\omega/c_{T}$, since the two halves of the radiation source
cancel at leading order. This means that the contribution of the
normal stress to the power radiated is higher order in $\omega$
for small $\omega$, and may be neglected.

The analysis proceeds as in the previous section. We again use
Eqs.~(\ref{in_plane_stress}) but now with $\Sigma_{t}=2\sqrt{2}E(I/w)k^{3}%
e^{-i\omega t}$ and $\Sigma_{n}\simeq0$. This gives the equations for the mode
amplitudes%
\begin{subequations}
\begin{align}
r^{2}(ik_{T}a_{T}+ik_{L}a_{L})+(r^{2}-2)(-ik_{T}a_{T}+i\zeta^{2}a_{L}/k_{L})
&  =0\\
i\zeta(a_{T}+a_{L})+(-ik_{T}^{2}a_{T}/\zeta+i\zeta a_{L})  &  =\Sigma
_{t}w/\bar{\mu}%
\end{align}
which can be solved to yield the response to leading order%
\end{subequations}
\begin{equation}
\frac{v(x=0)}{\Sigma_{t}}\simeq-\frac{r^{2}w}{\bar{\mu}}\frac{i}{\pi}\int
_{0}^{\infty}\frac{\kappa_{T}(\xi)}{F_{o}(\xi)}d\xi.
\end{equation}

(This result is analogous to Eq.~(124) of Miller and Pursey \cite{Miller54}).
This gives the average power radiated to leading order in small $\omega$%
\begin{equation}
P_{rad}=\frac{1}{2}w\operatorname{Re}(-\dot{v}\Sigma_{t}^{\ast})=\frac{1}%
{2}w(2\sqrt{2}EIk^{3}/w)^{2}\operatorname{Re}(i\omega v/\Sigma_{t})
\end{equation}
For the incident wave of unit amplitude we have
\begin{equation}
(u,u^{\prime},M,F)=(1,ik,EIk^{2},iEIk^{3})e^{i(kx-\omega t)}%
\end{equation}
so that the average incident power is%
\begin{equation}
P_{inc}=\frac{1}{2}\operatorname{Re}\left[  -i\omega(Fu^{\ast}+Mu^{\prime\ast
})\right]  =\omega EIk^{3}.
\end{equation}
The ratio gives the transmission coefficient%
\begin{equation}
\mathcal{T}_{o}(\omega)=\frac{1}{3}(wk)^{3}\left[  \frac{4}{\pi}%
\frac{(1+\sigma)}{(1-\sigma)}\operatorname{Re}\int_{0}^{\infty}\frac{\kappa
_{T}(\xi)}{F_{0}(\xi)}d\xi\right]  .
\end{equation}
Again the quantity in the braces turns out to be unity, so we have%
\begin{equation}
\mathcal{T}_{o}(\omega)=\frac{1}{3}(wk)^{3}.
\end{equation}

\subsubsection{Flexural Modes}

The flexural displacement $w(x,y)$ for a wave at frequency $\omega$ satisfies
the equation%
\begin{equation}
\nabla^{4}w=K^{4}w
\end{equation}
with $K^{2}=\sqrt{\rho d/D}\,\omega$, cf. section \ref{Subsec_Flexural}.
Expanding the cavity solution in transverse Fourier modes%
\begin{equation}
w(x,y)=\frac{1}{2\pi}\int_{-\infty}^{\infty}\tilde{w}(x,\zeta)e^{i\zeta
y}d\zeta
\end{equation}
the Fourier amplitude $\tilde{w}$ satisfies%
\begin{equation}
\left(  \partial^{2}/\partial x^{2}-\zeta^{2}\right)  ^{2}\tilde{w}%
=K^{4}\tilde{w}.
\end{equation}
The solutions are $\tilde{w}\sim e^{ikx}$ with%
\begin{equation}
k^{2}=\pm K^{2}-\zeta^{2}.
\end{equation}
Solutions corresponding to a wave propagating away from the source at $x=0$ or
exponentially decaying to $+\infty$ are given by $k_{\pm}$%
\begin{equation}
k_{+}=i\sqrt{K^{2}+\zeta^{2}}%
\end{equation}
and%
\begin{equation}
k_{-}=\left\{
\begin{tabular}
[c]{ll}%
$\sqrt{K^{2}-\zeta^{2}}$ & $\zeta^{2}<K^{2}$\\
$i\sqrt{\zeta^{2}-K^{2}}$ & $\zeta^{2}>K^{2}$%
\end{tabular}
\ \ \ \ \ \right.  .
\end{equation}
Thus the solution in the cavity can be written%
\begin{equation}
w(x,y)=\frac{1}{2\pi}\int_{-\infty}^{\infty}[\tilde{w}_{+}(\zeta)e^{ik_{+}%
x}+\tilde{w}_{-}(\zeta)e^{ik_{-}x}]e^{i\zeta y}d\zeta.
\end{equation}
The sources at $x=0$ are a moment $M(y)$ and an effective force per unit
length $V(y)$
\begin{subequations}
\label{Eq_FlexSources}%
\begin{align}
M(y)  &  =-D\left(  \frac{\partial^{2}w}{\partial x^{2}}+\sigma\frac{\partial
^{2}w}{\partial y^{2}}\right) \\
V(y)  &  =-D\left(  \frac{\partial^{3}w}{\partial x^{3}}+(2-\sigma
)\frac{\partial^{3}w}{\partial x\partial y^{2}}\right)  +w^{-1}(F_{c}%
(w/2)+F_{c}(-w/2))
\end{align}
with Fourier transforms $\tilde{M}(\zeta)$ and $\tilde{V}(\zeta)$. The last
two terms in the second equation are the corner forces. Matching these
boundary conditions gives
\end{subequations}
\begin{subequations}
\label{Eq_InPlaneSource}%
\begin{align}
\left[  K^{2}+\zeta^{2}(1-\sigma)\right]  \tilde{w}_{+}(\zeta)-\left[
K^{2}-\zeta^{2}(1-\sigma)\right]  \tilde{w}_{-}(\zeta)  &  =-\tilde{M}%
(\zeta)/D,\label{Eq_InPlaneMoment}\\
ik_{+}\left[  K^{2}-\zeta^{2}(1-\sigma)\right]  \tilde{w}_{+}(\zeta
)-ik_{-}\left[  K^{2}+\zeta^{2}(1-\sigma)\right]  \tilde{w}_{-}(\zeta)  &
=-\tilde{V}(\zeta)/D. \label{Eq_InPlaneForce}%
\end{align}

The average power radiated is%
\end{subequations}
\begin{equation}
P=-\left\langle \int\dot{w}(x=0,y)V(y)+\dot{\theta}%
(x=0,y)M(y)\,dy\right\rangle _{t}%
\end{equation}
with $\theta=-\partial w/\partial x$ the tilt angle and the dot denoting a
time derivative. For oscillations $e^{-i\omega t}$ the average over time gives%
\begin{equation}
P=P_{w}+P_{\theta}=-\frac{1}{2}\omega\operatorname{Im}\left[  \int
w(x=0,y)V^{\ast}(y)dy+\int\theta(x=0,y)M^{\ast}(y)\,dy\right]  .
\end{equation}
Evaluating the first integral in terms of Fourier expansions we find the power
radiated by the force%
\begin{equation}
P_{w}=-\frac{1}{4\pi}\omega\operatorname{Im}\left[  \int_{-\infty}^{\infty
}d\zeta\lbrack\tilde{w}_{+}(\zeta)+\tilde{w}_{-}(\zeta)]\tilde{V}^{\ast}%
(\zeta)\right]  . \label{Eq_Pw}%
\end{equation}
As in the scalar case, the imaginary part of this integral corresponds to the
excitation of propagating waves for which $\zeta<K$ and we may evaluate
$\tilde{V}(\zeta)$ to lowest nonzero order in $Kw$%
\begin{equation}
\tilde{V}(\zeta)=\int_{-w/2}^{w/2}dyV(y)e^{-i\zeta y}\simeq V_{0}-i\zeta
V_{1}+\cdots
\end{equation}
with%
\begin{align}
V_{0}  &  =\int dy\,V(y),\\
V_{1}  &  =\int dy\,yV(y).
\end{align}
We only keep the second term for antisymmetric sources for which $V_{0}$ is
zero. Note that $V_{0}$ is the total force normal to the plate, and $V_{1}$ is
the torque about the $x$ axis, and these can be evaluated from macroscopic arguments.

Similar arguments for $P_{\theta}$ give%
\begin{equation}
P_{\theta}=\frac{1}{4\pi}\omega\operatorname{Im}\left[  \int_{-\infty}%
^{\infty}d\zeta\lbrack ik_{+}\tilde{w}_{+}(\zeta)+ik_{-}\tilde{w}_{-}%
(\zeta)]\tilde{M}^{\ast}(\zeta)\right]
\end{equation}
with%
\begin{equation}
\tilde{M}(\zeta)\simeq M_{0}-i\zeta M_{1}+\cdots
\end{equation}
with $M_{0}$, $M_{1}$ the zeroth and first moments of $M$ over the bridge end.

Now we can calculate explicit results for the bending and torsion modes.

\paragraph{Bending mode}

From Eq.~(\ref{bend_junction_zeroth}) we see that an incident wave of unit
displacement amplitude $e^{i(kx-\omega t)}$ in the bridge gives oscillating
sources on the edge of the cavity%
\begin{align}
M_{0}  &  =2\sqrt{2}D(1-\sigma^{2})wk^{2}e^{i\pi/4}e^{-i\omega t},\\
V_{0}  &  =2\sqrt{2}D(1-\sigma^{2})wk^{3}e^{i\pi/4}e^{-i\omega t}.
\end{align}
Note that we are using the macroscopic formulation to derive these
expressions. It is somewhat subtle to directly use the expressions
Eq.~(\ref{Eq_FlexSources}) since the $y$ dependence of the mode
structure cannot be ignored. We have verified that these
expressions are reproduced using the long wavelength limit of the
mode structure given by solving Eqs.~(\ref{Eq_FlexEven},\ref{Eq_FlexOdd}).
Now defining $\xi=\zeta/K$ and%
\begin{equation}
\tilde{w}_{\pm}(\zeta)=-2\sqrt{2}we^{i\pi/4}u_{\pm}(\xi),
\end{equation}
using the dispersion relation $k^{2}=K^{2}/\sqrt{1-\sigma^{2}}$, and then
matching to the sources gives
\begin{subequations}
\label{Eq_upum}%
\begin{align}
\left[  1+\xi^{2}(1-\sigma)\right]  u_{+}-\left[  1-\xi^{2}(1-\sigma)\right]
u_{-}  &  =(1-\sigma^{2})^{1/2},\\
-\sqrt{1+\xi^{2}}\left[  1-\xi^{2}(1-\sigma)\right]  u_{+}-i\sqrt{1-\xi^{2}%
}\left[  1+\xi^{2}(1-\sigma)\right]  u_{-}  &  =(1-\sigma^{2})^{1/4}%
\end{align}
The power radiated is%
\end{subequations}
\begin{equation}
P=\omega Dw^{2}k^{2}K^{2}(1-\sigma^{2})\frac{2}{\pi}\operatorname{Im}\left[
\int_{-\infty}^{\infty}(u_{+}+u_{-})(1-\sigma^{2})^{-1/4}-(-\sqrt{1+\xi^{2}%
}u_{+}+i\sqrt{1-\xi^{2}}u_{-})\,d\xi\right]  .
\end{equation}
Normalizing by the incident power $P_{i}=\omega Dwk^{3}(1-\sigma^{2})$ gives
the transmission coefficient%
\begin{equation}
T_{0}(\omega)=kwI_{1}(\sigma),
\end{equation}
where $I_{1}$ is the integral%
\begin{equation}
I_{1}(\sigma)=\sqrt{1-\sigma^{2}}\frac{2}{\pi}\operatorname{Im}\int_{-\infty
}^{\infty}\left[  (u_{+}+u_{-})(1-\sigma^{2})^{-1/4}+(\sqrt{1+\xi^{2}}%
u_{+}-i\sqrt{1-\xi^{2}}u_{-})\,\right]  d\xi. \label{Eq_I1}%
\end{equation}
For $\sigma=0.33$, evaluating $u_{\pm}$ from Eq.\ (\ref{Eq_upum}) we find
\begin{equation}
I_{1}=2.3\quad(\sigma=0.33).
\end{equation}

\paragraph{Torsion Mode}

A unit amplitude mode $\theta=e^{i(kx-\omega t)}$ gives the oscillating torque
source%
\begin{equation}
\tau=4Dw(1-\sigma)ike^{-i\omega t}. \label{Eq_TMTorque}%
\end{equation}
Here the wave number in the beam is given by (cf. Eq.~(\ref{Eq_FlexTorsion}))%
\begin{equation}
kw=\frac{1}{2\sqrt{6(1-\sigma)}}(Kw)^{2}%
\end{equation}
with $K^{2}=\sqrt{\rho d/D}\omega$ as before. The torque
Eq.~(\ref{Eq_TMTorque}) corresponds to a force source term given
by a nonzero
first moment $V_{1}$of the tangential force%
\begin{equation}
V_{1}=\frac{2Dw^{2}K^{2}\sqrt{1-\sigma}}{\sqrt{6}}\text{.}%
\end{equation}
This gives a source term on the right hand side of
Eq.~(\ref{Eq_InPlaneForce}) $\tilde{V}(\zeta)=-i\zeta V_{1}$ and
the source term $\tilde{M}$ in Eq.~(\ref{Eq_InPlaneMoment}) is zero. Now defining%
\begin{equation}
\tilde{w}_{\pm}(\zeta)=\frac{2w^{2}\sqrt{1-\sigma}}{\sqrt{6}}u_{\pm}(\xi)
\end{equation}
with $\xi=\zeta/K$ from Eq.~(\ref{Eq_InPlaneSource}) we find
$u_{\pm}$ satisfies
\begin{subequations}
\label{Eq_upum_t}%
\begin{align}
\left[  1+\xi^{2}(1-\sigma)\right]  u_{+}-\left[  1-\xi^{2}(1-\sigma)\right]
u_{-}  &  =0,\\
-\sqrt{1+\xi^{2}}\left[  1-\xi^{2}(1-\sigma)\right]  u_{+}-i\sqrt{1-\xi^{2}%
}\left[  1+\xi^{2}(1-\sigma)\right]  u_{-}  &  =\xi.
\end{align}
Solving these equations for $u_{\pm}$ and so $\tilde{w}_{\pm}$,
substituting into Eq.~(\ref{Eq_Pw}) for the power radiated
($P_{\theta}$ does not contribute for this mode) and normalizing
by the incident power $\omega
Dw^{2}K^{2}\sqrt{1-\sigma}/2\sqrt{6}$ yields the transmission coefficient%
\end{subequations}
\begin{equation}
T_{0}(\omega)=kwI_{2}(\sigma)
\end{equation}
where $I_{2}$ is the integral%
\begin{equation}
I_{2}(\sigma)=(1-\sigma)\frac{4}{\pi}\operatorname{Im}\int_{-\infty}^{\infty
}(u_{+}+u_{-})\xi\,d\xi. \label{Eq_I2}%
\end{equation}
For $\sigma=0.33$, solving Eq.~(\ref{Eq_upum_t}) for $u_{\pm}$
yields
\begin{equation}
I_{2}=0.6\quad(\sigma=0.33).
\end{equation}

\section{Applications and Discussions}

\label{Sec_Applications}%

%TCIMACRO{\TeXButton{B}{\begin{table}[tbp] \centering}}%
%BeginExpansion
\begin{table}[tbp] \centering
%EndExpansion%
\begin{tabular}
[c]{||l||l|l|l|l||}\hline\hline
Mode & $\omega(k\rightarrow0)/c_{T}k$ & $\omega_{1}/(\pi c_{T}/w)$ &
$\mathcal{T}_{o}(k\rightarrow0)$ & $\mathcal{T}_{o}(\omega\rightarrow
0),\,z=\omega w/c_{T}$\\\hline\hline
Compression & $\sqrt{2(1+\sigma)}$ & $\sqrt{\frac{2}{1-\sigma}}$ & $4kw$ &
$2\sqrt{\frac{2}{1+\sigma}}z$\\\hline
Torsion & $2d/w$ & $\frac{8.014}{\sqrt{(1-\sigma)}}(\frac{d}{w})$ & $I_{2}kw$
& $\frac{1}{2}I_{2}\left(  \frac{w}{d}\right)  z$\\\hline
In-plane bend & $\sqrt{\frac{1+\sigma}{6}}kw$ & $1$ & $\frac{1}{3}(kw)^{3}$ &
$\frac{1}{3}\left(  \frac{6}{1+\sigma}\right)  ^{3/4}z^{3/2}$\\\hline
Flex-bend & $\sqrt{\frac{1+\sigma}{6}}kd$ & $\frac{2.886}{\sqrt{(1-\sigma)}%
}(\frac{d}{w})$ & $I_{1}kw$ & $I_{1}\left(  \frac{6}{1+\sigma}\right)
^{1/4}\left(  \frac{w}{d}\right)  ^{1/2}z^{1/2}$\\\hline\hline
\end{tabular}
\caption{
Long wavelength properties of the modes of a thin beam. See text for the details.
\label{Table1}%
}
%TCIMACRO{\TeXButton{E}{\end{table}}}%
%BeginExpansion
\end{table}%
%EndExpansion

The results for the long wavelength properties a thin plate beam
of width $w$ and thickness $d$ are brought together in table
\ref{Table1}. The second column gives the small $k$ dispersion
relation in terms of the propagation speed $c_{T}$ of the in-plane
shear wave in a thin plate which is the same as the shear wave
speed in the bulk medium. The third column gives the frequency
cutoff $\omega_{1}$ of the lowest waveguide mode with the same
transverse parity symmetry as the acoustic mode (this would be
$2\Delta$ in the scalar model). The fourth column expresses the
small $\omega,k$ energy transmission coefficient $\mathcal{T}_{o}$
of the acoustic mode in terms of the wave number $k$ in the bridge
and the width $w$ of the bridge. This is useful in considering the
$Q$ of the fundamental vibration modes of the beam for which $k$
is of order $\pi/L$ with $L$ the length of the beam. The
quantities $I_{1}$ and $I_{2}$ are Poisson ratio dependent numbers
defined by Eqs.~(\ref{Eq_I1},\ref{Eq_I2}) and take on the values
$I_{1}=2.3$ and $I_{2}=0.6$ for $\sigma=1/3$. Finally, the fifth
column re-expresses the small $\omega,k$ dependence of the energy
transmission coefficient $\mathcal{T}_{o}$ in terms of the
frequency $\omega$. This form is particularly useful to estimate
the reduction from the universal thermal conductance at low
temperatures due to the strong scattering of the long wavelength
modes by an abrupt junction.

\subsection{Implications for Heat Transport}

The low frequency asymptotic limit of the transmission coefficient
$\mathcal{T}_{o}(\omega\rightarrow0)$ allows us to calculate the low
temperature variation of the thermal conductivity. If we write the dependence
as%
\begin{equation}
\mathcal{T}_{o}(\omega\rightarrow0)=A(\omega w/c_{T})^{p}%
\end{equation}
then relative to the universal low temperature one mode value $\left.
K/T\right|  _{u}=\pi^{2}k_{B}^{2}/3h$ we have for each acoustic mode
\begin{equation}
\frac{K/T}{\left.  K/T\right|  _{u}}=A\left(  \frac{k_{B}Tw}{\hbar c_{T}%
}\right)  ^{p}\frac{3}{\pi^{2}}\int_{0}^{\infty}\frac{x^{2+p}e^{x}}%
{(e^{x}-1)^{2}}\,dx,
\end{equation}
where the prefactor $A$ for each mode can be found from table \ref{Table1}.
The integral is just some $p$-dependent constant. Using the expressions in
table \ref{Table1} the result can be written in terms of the frequency
$\omega_{1}$ cutoff of the first wave guide mode of the corresponding symmetry%
\begin{equation}
\frac{K/T}{\left.  K/T\right|  _{u}}=B\left(  \frac{k_{B}T}{\hbar\omega_{1}%
}\right)  ^{p}.
\end{equation}
The prefactor $B$ is a numerical constant that depends on the
Poisson ratio $\sigma$, but \emph{not} on geometrical factors such
as $d/w$. Since the thermal excitation of the waveguide modes
occurs for $k_{B}T\gtrsim 0.2\times\hbar\omega_{1}$ (cf.
Fig.~\ref{Fig_K_Tnew}) this expression indicates to what degree
the plateau in $K/T$ becomes apparent as the temperature is
lowered and the waveguide mode freezes out, before the reduced
transmission coefficient at small frequencies begins to lower
$K/T$ to zero. The ideal low temperature universal value of $K/T$
will be more evident for smaller powers $p$ (cf. the comparison of
the two scalar results for $p=1$ and $p=3$ in
Fig.~(\ref{Fig_K_Tnew})). This suggests that the compression and
torsion modes will give contributions to $K/T$ curves similar to
the result for the stress free scalar model in
Fig.~\ref{Fig_K_Tnew}, without a well defined plateau at low
temperatures (all have $p=1$), and the in-plane bend mode
($p=3/2$) will probably have no indication of a plateau. On the
other hand for the flexural bend mode ($p=1/2$) the transmission
coefficient increases more rapidly with increasing frequency
$\mathcal{T}_{o}\sim \sqrt{\omega/\omega_{1}}$. This will lead to
a more rapid increase in $K/T$ towards the universal value before
significant excitation of additional modes occurs at
$T\sim\hbar\omega_{1}/k_{B}$, leading to a more pronounced
plateau. It should be noted that for this mode the plateau in
$K/T$ only develops at very low temperatures in the thin plate
limit, reduced from the simple estimate $\hbar c_{T}/k_{B}w$ by
the ratio $d/w$ of the thickness to the width of the beam.

\subsection{Implications for Q}

Using the third column of the table and Eq.~(\ref{Q_transmission}) for $Q$ we
get the simple estimates $Q\sim L/w$ for the fundamental compression mode,
torsion mode, and flexural-bending mode, and $Q\sim(L/w)^{3}$ for the in-plane
bending mode. Only for the in-plane bending mode is the isolation of the
bridge modes from the supports sufficiently strong to give a large $Q$ for
accessible geometries (e.g. $L/w<100$). The strong isolation of this mode is
easy to understand physically: the wide supports are very rigid against
bending motion in the plane.

In many experiments on mesoscopic oscillators, modes other than the in-plane
bending modes are used, and values of $Q$ significantly higher than the value
suggested by the geometric ration $L/w$ are obtained. One way this is done is
to use more complicated geometries, such as compound torsional oscillators
arranged so that the amplitude of vibration in the bridge supports is reduced.
Also, in oscillators at larger scales it is relatively easy to produce more
rigid supports, for example by fabricating a bridge or cantilever making an
abrupt junction to a three dimensional support, which can also be of a
different elastic material, both of which will reduce the coupling to the
support modes. In mesoscopic oscillators, where the geometry is typically
etched out of a single material, and undercutting by the etch comprises
attempts to make an abrupt junction to a three dimensional support, our
estimates of the coupling will be more appropriate.

Our estimates of $Q$ suppose that all the energy communicated to the support
modes is lost from the energy of the oscillator. This is not necessarily the
case, for example if the support material is also of sufficiently low loss and
isolated from the rest of the experiment. However, our results do show that
when the bridge-support coupling is large, it is important to consider the
dissipation properties of the support structures as well as the bridge,
cantilever, or other oscillator that is the obvious focus of attention.

\section*{Acknowledgments}

This work was supported by the National Science Foundation under
grant number DMR-9873573 and DARPA MTO/MEMS under Grant No.
DABT63-98-0012. We thank Michael Roukes, Andrew Cleland, and Keith
Schwab for many useful discussions, and Deborah Santamore for
carefully reading the manuscript.

%\bibliographystyle{prsty}
%\bibliography{all,phonons}

\end{document}